\documentclass[preprint2]{aastex62}
\usepackage[T1]{fontenc}
\usepackage{bm}        
\usepackage{amssymb, amsmath}   
\usepackage{cancel}
\usepackage{color}
\usepackage{placeins}

\DeclareMathAlphabet{\mathsfit}{\encodingdefault}{\sfdefault}{m}{sl}
\SetMathAlphabet{\mathsfit}{bold}{\encodingdefault}{\sfdefault}{bx}{sl}

\DeclareMathOperator{\sech}{sech}

\received{January 8, 2020}
\revised{****}
\accepted{****}
\submitjournal{ApJ}
\shortauthors{P. Kilian, X. Li, F. Guo and H. Li}

\begin{document}

\title{Exploring the acceleration mechanisms for particle injection and power-law formation during trans-relativistic magnetic reconnection}
\shorttitle{Exploring the acceleration mechanisms in trans-relativistic reconnection}
\correspondingauthor{Patrick Kilian}
\email{pkilian@lanl.gov}

\author[0000-0002-8906-7783]{Patrick Kilian}
\affil{Los Alamos National Laboratory, Los Alamos, NM 87545}
\affil{Centre for Space Research, North-West University, Potchefstroom, South Africa}
\author[0000-0001-5278-8029]{Xiaocan Li}
\affil{Dartmouth College, Hanover, NH 03750 USA}
\author[0000-0001-5278-8029]{Fan Guo}
\affil{Los Alamos National Laboratory, Los Alamos, NM 87545}
\affil{New Mexico Consortium, Los Alamos, NM 87544}
\author[0000-0003-3556-6568]{Hui Li}
\affil{Los Alamos National Laboratory, Los Alamos, NM 87545}

\begin{abstract}
Magnetic reconnection in the relativistic and trans-relativistic regimes is able to accelerate particles to hard power law energy spectra $f \propto \gamma^{-p}$ (approaching $p=1$). The underlying acceleration mechanism that determines the spectral shape is currently a topic of intense investigation. By means of fully kinetic plasma simulations, we carry out a study of particle acceleration during magnetic reconnection in the trans-relativistic regime of a proton-electron plasma. While earlier work in this parameter regime has focused on the effects of electric field parallel to the local magnetic field on the particle injection (from thermal energy to the lower energy bound of the power-law spectrum), here we examine the roles of both parallel and perpendicular electric fields to gain a more complete understanding on the injection process and further development of a power-law spectrum. We show that the parallel electric field does contribute significantly to particle injection, and is more important in the initial phase of magnetic reconnection. However, as the simulation proceeds, the acceleration by the perpendicular electric field  becomes more important for particle injection and completely dominates the acceleration responsible for the high-energy power-law spectrum. This holds robustly, in particular for longer reconnection times and larger systems, i.e. in simulations that are more indicative of the processes in astrophysical sources.
\end{abstract}


\section{Introduction}
\label{sec:intro}

Magnetic reconnection is the process that changes magnetic field topology.
This requires at least a local violation of the frozen-flux theorem.
In typical astrophysical, collisionless plasmas this happens due to kinetic, micro-physical processes.
Once the magnetic field lines are broken and reconnected, the magnetic field in its new topology relaxes to a lower energy configuration at much larger scales.
The released energy in converted into heating, bulk flows and a tail of high-energy particles \citep{Zelenyi_1990,Zenitani_2001, Guo_2014}. 
This source of non-thermal particles is thought to be important in a number of high-energy astrophysical environments such as pulsar wind nebulae, gamma-ray bursts, and jets from active galactic nuclei.
Knowing the process of particle acceleration is important for making predictions of the particle energy spectra.

Past research on particle acceleration during magnetic reconnection has mainly explored two mechanisms: 1. Fermi-type acceleration where particles are accelerated by bouncing back and forth in the reconnection generated flows \citep{Dal_Pino_2005,Drake_2006,Fu_2006,Drury_2012,Guo_2014,Dahlin_2014,LeRoux_2015,Li_2018a,Li_2018b,Li_2019} and 2. direct acceleration at diffusion regions surrounding reconnection X-points \citep{Zenitani_2001,Pritchett_2006,Cerutti_2013,Wang_2016,Sironi_2014}. The Fermi-type acceleration is mainly through the electric field induced by bulk plasma motion $\vec{E}_m = - \vec{u} \times \vec{B}/c$ perpendicular to local magnetic field, whereas the direct acceleration is driven by the parallel electric field if a non-zero magnetic field exists. It is therefore useful to distinguish the relative contribution of the two during the particle acceleration process, either according to the generalized Ohm's law \citep{Guo_2019}, or simply by decomposing the electric field into the perpendicular part $E_\perp$ and parallel component $E_\parallel$ and evaluate the work done by each of them \citep{Guo_2015,Ball_2019}.

Since the magnetic field is the source of free energy for these energization processes, it is useful to define two parameters that compare the magnetic field with other characteristic properties of the plasma.
The first parameter is the magnetization $\sigma = B_0^2 / (4 \pi \rho \mathrm{c}^2)$ where $B_0$ is the magnetic field strength, $\rho$ is mass density and $\mathrm{c}$ is the speed of light. This ratio between energy density in the magnetic field to the energy density associated with the rest mass of the particles can also be seen as the energy available per particle from the magnetic field if the magnetic energy was fully converted through reconnection.
The second parameter is the plasma $\beta$ defined as $\beta = 8 \pi n k_B T /B_0^2$. This ratio compares the thermal pressure of the gas with the pressure due to the magnetic field.
Alternatively this can be expressed by $\sigma_\mathrm{th} = B_0^2/(12 \pi n k_B T) = 2/(3\,\beta)$, the ratio between magnetic field energy and thermal energy density that measures the maximum possible energization per particle compare with the thermal energy.

Magnetic reconnection is especially interesting in the case of $\beta \lesssim 1$ since more energy is available for particle energization.
The non-relativistic case of $\sigma \ll 1$ of magnetic reconnection has been studied for many years \citep{Zelenyi_1990,Biskamp_1996,Birn_2001,Hesse_2001,Priest_2007,Shay_2007,Treumann_2013,Munoz_2015,Li_2018a,Li_2019}. This parameter regime is especially relevant to space and solar physics and is accessible in laboratory experiments. 
More recently (ultra-)relativistic reconnection with $\sigma \gg 1$ has also been studied \citep{Zenitani_2001,Lyubarsky_2008,Sironi_2014,Guo_2014,Guo_2015,Guo_2016,Werner_2015,Liu_2015,Liu_2017,Liu_2019}.
This parameter regime is of particular interest in astrophysics, such as jets from active galactic nuclei or pulsar magnetospheres. In these systems $\sigma$ is so large that, even if only a small fraction of the magnetic energy is released, particles can still reach relativistic energies. Simulations show that the resulting particle distribution function often have hard power law tails that extend to large Lorentz factors \citep{Sironi_2014,Guo_2014,Guo_2015,Guo_2016,Werner_2015}. These particle spectra might be responsible for observed high-energy emission through inverse-Compton up-scatter of softer seed photons, synchrotron radiation from the gyration of the energetic particles in the magnetic field, or other processes such as Bremsstrahlung.

In recent years the trans-relativistic regime $\sigma \approx 1$ has generated interest as well \citep{Melzani_2014a, Melzani_2014b, Rowan_2017, Werner_2017, Ball_2018, Rowan_2019, Ball_2019}. One particularity of this regime is that the the available energy compared to the rest mass of ions $\sigma_i = B_0^2 / (4 \pi \rho_i \mathrm{c}^2)$ is around unity ($\sigma_i \approx 1$, where $\rho_i = m_i n_i$ is the mass density of ions with mass per ion of $m_i$ and number density $n_i$). For electrons with mass $m_e$ and number density $n_e \approx n_i$ however $\sigma_e = B_0^2 / (4 \pi \rho_e \mathrm{c}^2) \gg 1$. When magnetic field energy is converted, this leads to only mildly relativistic ions but a fraction of electrons can be very relativistic. 

One astrophysical class of objects where $\sigma \approx 1$, $\beta \lesssim 1$ is thought to occur are radiatively inefficient, geometrically thick, optically thin accretion discs discs around black holes that accrete much less then their Eddington limit. Even in disc that accrete more rapidly this condition might be satisfied in the corona above the disk \citep{DiMatteo_1998}. 
Simulation of magneto-hydrodynamics that include effects of general relativity (GRMHD) give a sense of the overall flow geometry and energy contained in radiation, magnetic field and ion fluid \citep{Chan_2015,Porth_2017,Davelaar_2018,Chael_2019,Mahlmann2020,Nathanail2020}. The energy content in the electron is basically unconstrained by single fluid simulations, but is of importance for predictions of observable electromagnetic signatures since the electrons radiate away their energy much more readily. It is therefore interesting to study the process of electron energization in much more detail.

The work by \cite{Rowan_2017} concentrated on the heating of the electrons and ions and the temperature ratio $T_e/T_i$. They also investigated the artificial influence of the numerical mass ratio $m_i / m_e$. Simulations with full mass ratio are possible and preferable, as simulations with too small mass ratio can overestimate the heating rate.
Above the energies found in the heated distribution, the particle spectrum tends to form a power law with an exponential cut-off $f(\gamma) \propto \gamma^{-p} \exp{(\gamma / \gamma_c)}$. In \cite{Werner_2017} the scaling of $p$ and $\gamma_c$ with magnetization $\sigma$ is studied in the range $\sigma_i = 0.03 \dots 10^4$ and an empirical fit formula is provided. The authors also investigate the energy partition between electrons and protons. They find that towards the non-relativistic regime electrons only receive about 1/4 of the energy, but electrons and protons obtain equal amount of energy as $\sigma_i$ gets large.

In \cite{Ball_2018} a similar study is performed for $\sigma = 0.3 \dots 3$, but additionally the dependence on $\beta$ in the range $10^{-4}\dots 1.5$ is investigated.
The paper also investigates the role of X-point acceleration and Fermi-acceleration and presents some resulting particle trajectories and spectra.
 \citet{Ball_2019} continues the particle acceleration study and focuses on the question how particles get into the power-law tail of the energy distribution. To this end the authors select a Lorentz factor $\gamma_{inj} = \sigma_e /2$ that separates the low energy part of the energy distribution from the power-law tail in their simulations with $\sigma_i = 0.3$. The analysis is mostly limited to the role of $W_\parallel$ and how and where particles cross $\gamma_{inj}$.

In this paper, we extend this analysis and investigate the role of both the parallel and perpendicular electric fields where the direction are defined with respect to the local magnetic field. To ensure that our results that can be readily compared with  \citet{Ball_2019} we chose identical parameters wherever feasible. The only difference is the choice of initial equilibrium (we chose a force-free current sheet instead of a Harris sheet) and the form of initial perturbation (we perturb the magnetic field instead of a localized reduction in thermal pressure). However, we believe the two simulations are sufficiently similar for comparison. We do however limit our investigation to the triggered case and only consider $B_g = 0.1 B_0$, omitting the case of $B_g = 0.3 B_0$. Effects of guide field will be discussed in a forthcoming study. We find that parallel electric field does accelerate a fraction of electrons to the injection energy in the initial phase of reconnection. However, as reconnection proceeds, the acceleration by perpendicular electric field becomes important and outperform the parallel electric field for particle injection. Moreover, the acceleration from the injection energy into power law energies is completely dominated by perpendicular electric field. The resulting power-law index does not seem to strongly depend on the injection mechanism. These provide further evidence for the roles of the Fermi-like process in determining particle acceleration into a power-law spectrum during magnetic reconnection.


The remainder of the paper is organized as follows: In the following Sec.~\ref{sec:setup} we describe the simulation code and the initial setup which results in magnetic reconnection in the desired trans-relativistic regime.
In Sec.~\ref{sec:results} the results obtained from the simulation are described.
In Sec.~\ref{sec:conclusion} we discuss and draw conclusions on the relative roles of the parallel and perpendicular electric field.

\section{Simulation Setup}
\label{sec:setup}

\begin{figure}[htb]
    \includegraphics[width=\columnwidth]{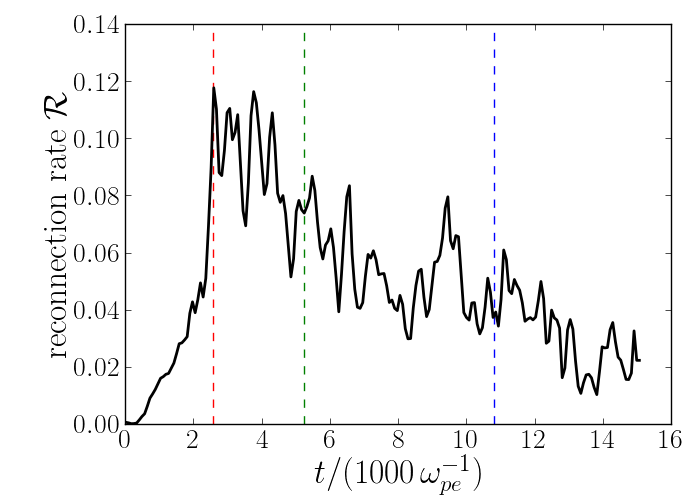}\\
    \includegraphics[width=\columnwidth]{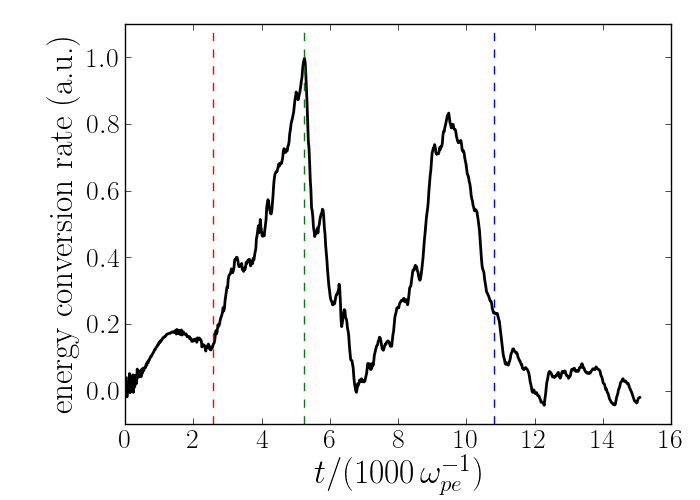}
    \caption{Top: Reconnection rate $R = (c\,E_\mathrm{rec})/(v_A \, B_0)$  based on the reconnection electric field $E_\mathrm{rec}$. The vertical lines indicate the three points in time (peak of reconnection rate, peak of energy conversion rate and reconnection close to saturation) for further analysis. Bottom: rate of magnetic energy conversion in arbitrary units.}
    \label{fig:reconnectionrate}
\end{figure}

\begin{figure}[thb]
    \includegraphics[width=\columnwidth]{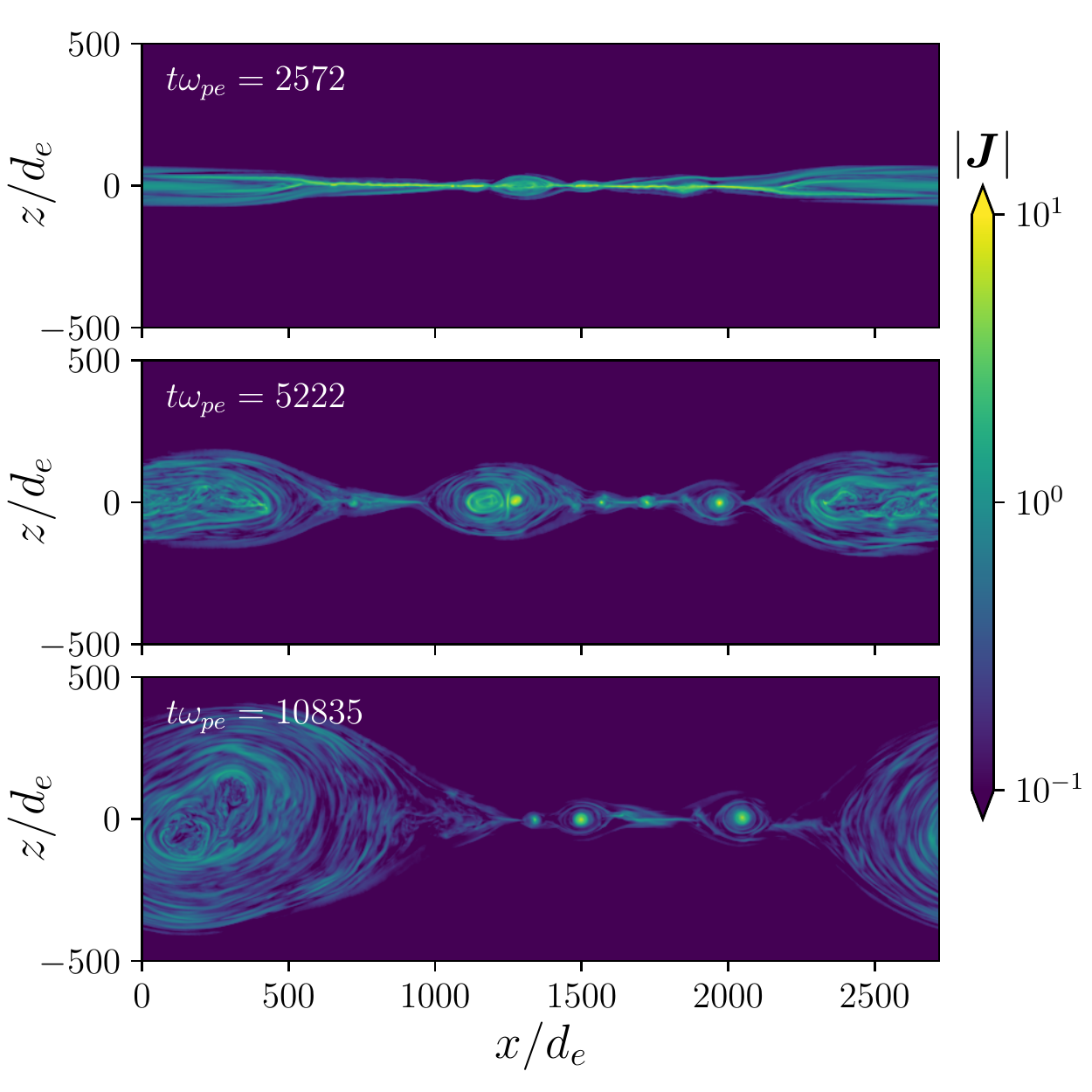}
    \caption{Magnitude of the current density. The three points in time correspond (as closely as output settings allow) to the points in time that are highlighted in Fig.~\ref{fig:reconnectionrate}.
    \label{fig:absJ}}
\end{figure} 

To study the processes involved in electron energization we perform fully-kinetic simulations using the VPIC code by \cite{Bowers_2008a, Bowers_2008b, Bowers_2009}. The code uses a relativistically correct implementation of the Boris push \citep{Boris_1970} to move macro-particles that represent phase space density and interpolates to and from the Eulerian grid using a low-order energy-conserving scheme. The current deposition is constructed such that the continuity equation between particles charges and currents on the grid is satisfied. Additionally, deviations from Gauss's Law are cleaned periodically to prevent round-off errors from accumulating. The electromagnetic fields on the grid are advanced using a standard Yee scheme \citep{Yee_1966}.

Inside the code we set up a single force-free current sheet in a electron-proton plasma with natural mass ratio $m_i / m_e = 1836$. We choose a force-free current sheet instead of a Harris current sheet as this removes the arbitrary choice of over-density $\eta$ inside the current sheet compared to the upstream plasma. To remove any influence of this choice on the analysis results, we show results including and excluding the particles that start inside the current sheet in the analysis below. Our two-dimensional simulation resolves the $x$ and $z$ directions, where the $x$ direction is along the upstream magnetic field and the magnetic field direction varies with the $z$ direction across the current sheet. The $x$ direction has periodic boundary conditions on particles and fields. The $z$ direction is terminated by fixed walls at the top and bottom of the domain. These walls are electrically perfectly \textbf{conducting} \sout{conduction} and flip the $v_z$ velocity component of particles hitting the wall. This way the $z$ boundaries reflect electromagnetic waves as well as particles. There is no inflow of plasma or magnetic flux from the $z$ direction and the reconnection exhaust eventually interacts with each self across the periodic $x$ boundary. This terminates reconnection eventually as discussed below. We do include a guide field $B_g$ that is perpendicular to the anti-parallel reconnecting magnetic field $B_0$. In the upstream this field is in the out of plane direction and has strength $B_g = 0.1 B_0$.
The initial field configuration is given by
\begin{align}
  B_x &= B_0\,\tanh{(z / \lambda)} + \delta B_x\quad ,\\
  B_y &= B_0\,\sqrt{(B_g/B_0)^2 + \sech^2(z / \lambda)} \quad , \\
  B_z &= \delta B_z \quad .
\end{align}
This magnetic field has constant magnitude $\sqrt{B_0^2 + B_g^2}$ and rotates through an angle $\pi - 2 \arctan{B_g/B_0} \approx 169^{\circ}$ when crossing the current sheet in the $z$ direction. In terms of the electron skin depth $d_e = c / w_{pe}$ the characteristic thickness of the initial current sheet is $\lambda = 14.1\, d_e$. Reconnection is triggered by a divergence-free, long wavelength perturbation that is described by
\begin{align*}
    \delta B_x &= -  \frac{L_x B_0}{100L_z} \cos{\left(\frac{2\pi\,(x-0.5L_x)}{L_x}\right)} \sin{\left(\frac{\pi\,z}{L_z}\right)}  , \\
    \delta B_z &= \phantom{-} \frac{B_0}{50} \sin{\left(\frac{2\pi\,(x-0.5L_x)}{L_x}\right)} \cos{\left(\frac{\pi\,z}{L_z}\right)} .
\end{align*}

We primarily discuss the case where the size of the simulation domain is $L_x \times L_y \times L_z = 2720\,d_e \times  0.33\,d_e \times 1360\,d_e$ and is resolved by $N_x \times N_y \times N_z = 8192 \times 1 \times 4096$ grid cells. In addition, we have explored effects of the box sizes and results are summarized in the Appendix~A.
The time step was set to $\Delta{t} = 0.188 \omega_{pe}^{-1}$, 80 percent of the maximum time step permitted by the CFL condition.
Ions and electrons are each represented by 100 particles per cell.
Our choice of $\sigma_i = 0.3$ sets the plasma density and implies $\sigma_e \approx 552.5$. Together with the mass ratio this also  sets $\omega_{pe} / \Omega_\mathrm{ci} = 78.1$. 
The initial plasma temperature is given by $k_B\,T_e = k_B\,T_i = 0.918\,m_e\,c^2$. The plasma beta resulting from these parameters is $\beta = 3.3\cdot10^{-3}$, $\sigma_\mathrm{th} = B_0^2/(12 \pi n k_B T) = 2/(3\,\beta) \approx 200$.

To analyze the particle acceleration process, a small fraction of the macro-particles -- 0.67 million electrons -- are randomly selected in the beginning of the simulation and designated as tracer particles. For these particles quantities of interest such as the work done by the electric field and the parallel and perpendicular components $E_\parallel$ and $E_\perp$ is computed in every time step and is output for later analysis.

\section{Results}
\label{sec:results}

As expected the perturbed equilibrium is unstable and quickly starts to reconnect at the induced X-point and secondary X-points that form between plasmoids in the collapsing current sheet \citep{Liu_2020}. A time history of the reconnection rate $R = (c\,E_\mathrm{rec})/(v_A \, B_0)$ based on the reconnection electric field $E_\mathrm{rec}$ is plotted in Fig.~\ref{fig:reconnectionrate}. $E_\mathrm{rec}$ is computed from the time derivative of the magnetic flux $\Psi$, which in turn is obtained by calculating the out-of-plane component of the vector potential $A_\mathrm{y}$ from $B_\mathrm{x}$ and $B_\mathrm{z}$ and taking $\Psi = \max{(A_\mathrm{y})} - \min{(A_\mathrm{y})}$. 
Taking the inflow speed $v_\mathrm{in}$ and the outflow speed $v_\mathrm{out} \approx v_A$ we find that the reconnection rate is similar to their ratio, i.e., $R \approx v_\mathrm{in}/v_A$.

Based on the reconnection rate we choose the three points in time in Fig.~\ref{fig:reconnectionrate} for the analysis presented later in the paper. Fig.~\ref{fig:absJ} shows the distribution of current density at the three times.
The first point in time is $t_1 = 2572 \omega_{pe}^{-1}$.
At this time three islands  and reconnection outflows from the X points between them have formed and the reconnection rate peaks. The second point in time that we chose is $t_2 = 5259 \omega_{pe}^{-1}$
at the peak of the magnetic energy conversion rate (lower panel in Fig.~\ref{fig:reconnectionrate}). Several islands of different sizes and different spatial separation in the $x$ direction exist and two plasmoids are merging in the middle of the domain, driving a spike in the energy conversion rate as shown in the lower panel in Fig.~\ref{fig:reconnectionrate}. As for the third time step we pick $t_3 = 10819 \omega_{pe}^{-1}$
, which is just short of two Alfv\'en crossing times. At this point the energy conversion rate and reconnection rate have  dropped significantly and the self-interaction of the system via the periodic boundary start to be important.

\begin{figure*}[htb]
    \begin{center}
    \includegraphics[width=\textwidth]{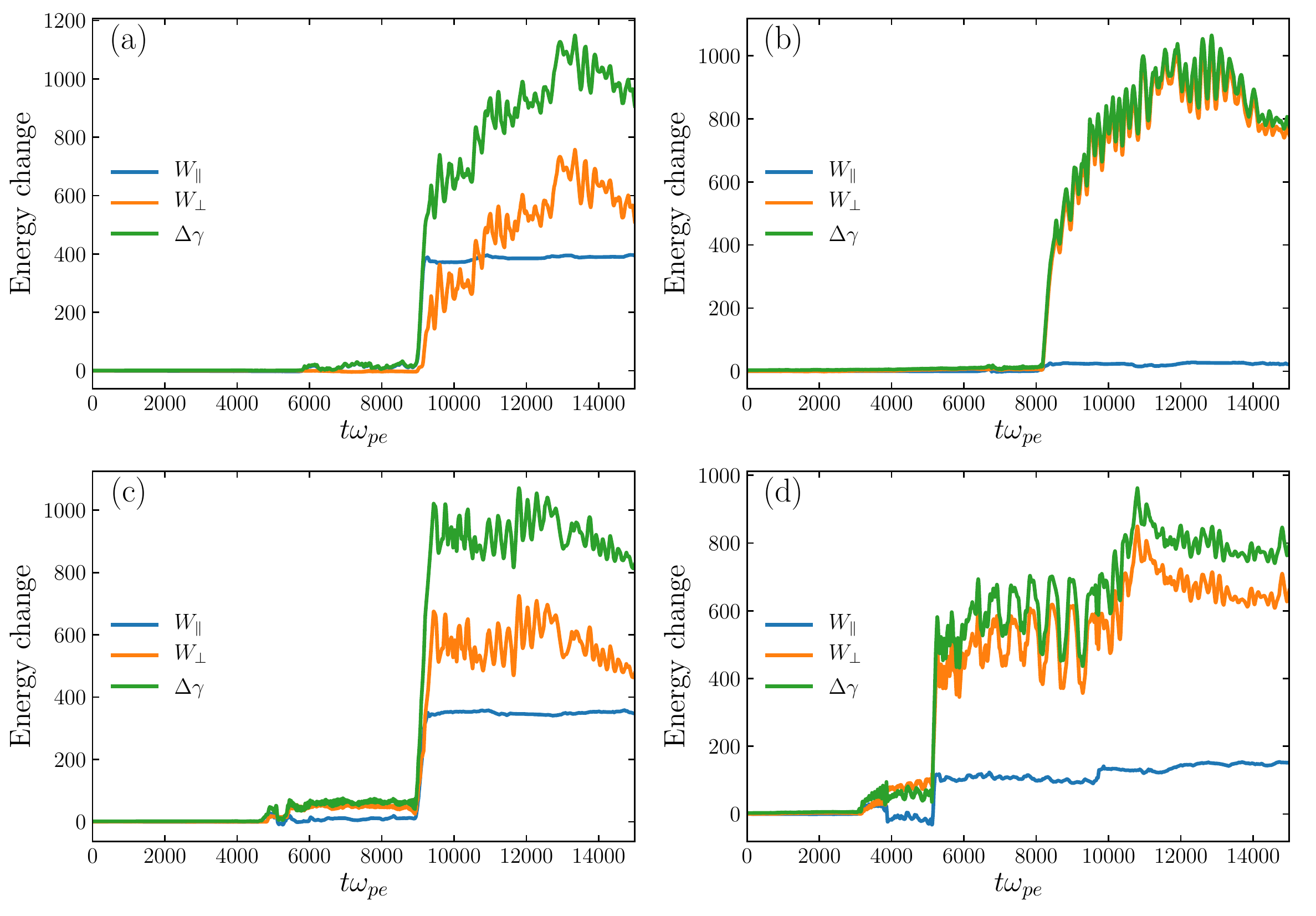}
    \end{center}
    
    \caption{Four electron trajectories. a) The top left trajectory illustrates the case where a particle is first injected due to $W_\parallel$ and then gains further energy due to $W_\perp$. b) The second trajectory illustrates that $W_\perp$ can directly inject particles with negligible contribution from $W_\parallel$. c) The third trajectory illustrates that both components of the electric field can act simultaneously during injection. d) The fourth trajectory illustrates that for many particles classification of the injection mechanism is not straight forward.}
    \label{fig:traj}
\end{figure*}

As mentioned before we employ a reduced, but statistically significant number of tracer particles. This allows output at high cadence including diagnostic quantities that are computed in every time step. The most important quantities used in this paper are the work done by the parallel and perpendicular electric fields, where the direction are split with respect to the local magnetic field at the particle location. The full definition of the two quantities is
\begin{align}
    W_\parallel(t) &= \frac{q}{m_e\,c^2}\int_0^t \vec{v}_p(t') \cdot \vec{E}_\parallel(t')\,\mathrm{d}t'  \\ 
    W_\perp(t) &= \frac{q}{m_e\,c^2}\int_0^t \vec{v}_p(t') \cdot \vec{E}_\perp(t')\,\mathrm{d}t'  \\
    \mathrm{where}& \nonumber \\
    \vec{E}_\parallel(t') &= \frac{\vec{E}(t')\cdot\vec{B}(t')}{\vec{B}(t')\cdot\vec{B}(t')}\vec{B}(t') \\
    \vec{E}_\perp(t') &= \vec{E}(t') - \vec{E}_\parallel(t')
\end{align}
In this definition we use the electric field $\vec{E}(t)$ at time $t$ at the particle location $\vec{x}_p(t)$, the magnetic field $\vec{B}(t)$ at the same time and location, the (negative) electron charge $q$, electron rest mass $m_e$ and the electron velocity $\vec{v}_p(t)$. 
Since we have a small guide field that avoids magnetic nulls, the decomposition parallel and perpendicular to the magnetic field is meaningful nearly everywhere in the simulation box. The parallel component can only be generated by non-ideal, kinetic processes, whereas the perpendicular component can be generated by bulk plasma motion $- \vec{u}\times\vec{B}/c$ or the Hall term. The perpendicular component therefore does not require kinetic, non-ideal effects. While these non-ideal effects can contribute to the perpendicular component, we do not attempt to split the perpendicular component into ideal and non-ideal components in this work. Instead, we note that any perpendicular electric field can support Fermi acceleration in a general sense \citep{Lemoine_2019}. This is in line with the conclusion that the Fermi acceleration in reconnection is driven by particle curvature drift motions along the perpendicular electric field \citep{Guo_2014,Dahlin_2014,Li_2017}.

Fig.~\ref{fig:traj} shows the time history of $W_\parallel$, $W_\perp$ and $\Delta \gamma$ for four representative tracer particles. The first trajectory (panel a) shows a particle that rapidly gains energy to $\gamma \approx 400$ through parallel electric field at time $t \approx 8800 \omega_{pe}^{-1}$. After this single short episode the particle gains more energy through the perpendicular field over the next $4000 \omega_{pe}^{-1}$. This particle trajectory provides evidence that the parallel electric field can provide an ``injection'' process for further energization through $E_\perp$, similar to \citep{Ball_2018, Ball_2019}. 
However, inspecting trajectories of the $500$ most energetic tracer particles also reveals other acceleration patterns, which are shown in the same figure. The top right trajectory (panel b) shows a particle that never gained appreciable energy from $W_\parallel$, but was picked up by $W_\perp$ at $t \approx 8100 \omega_{pe}^{-1}$ and kept gaining energy more slowly after $\gamma \approx 400$ \citep{Guo_2015,Sironi_2019}. The duration of injection, subsequent acceleration, and final particle energy are very comparable to the previous trajectory, but without appreciable input from $W_\parallel$. It is worth noting that the time of injection is also similar to the first particle and coincided with the merger of two large islands. 

We also find several trajectories similar to the one displayed in the bottom left corner (panel c), where $W_\parallel$ and $W_\perp$ have remarkably comparable contributions during the injection. In addition, there are many trajectories that defy simple classification such as the one shown in the bottom right (panel d) where $W_\parallel$ changes sign several times (note that this is integrated work done, not instantaneous power). We also find many trajectories where there is no clear injection moment, but rather a gradual increase in energy over a period of several thousand $\omega_{pe}^{-1}$ through a Fermi-like process. This may have important implications, as it suggests that Fermi acceleration does not depend on a real injection process in magnetic reconnection.  These trajectories show that the injection process is more complicated than what is shown in \citet{Ball_2018,Ball_2019}. Therefore more careful future studies on the low-energy acceleration process is desired.

\begin{figure}[htb]
    \includegraphics[width=\columnwidth]{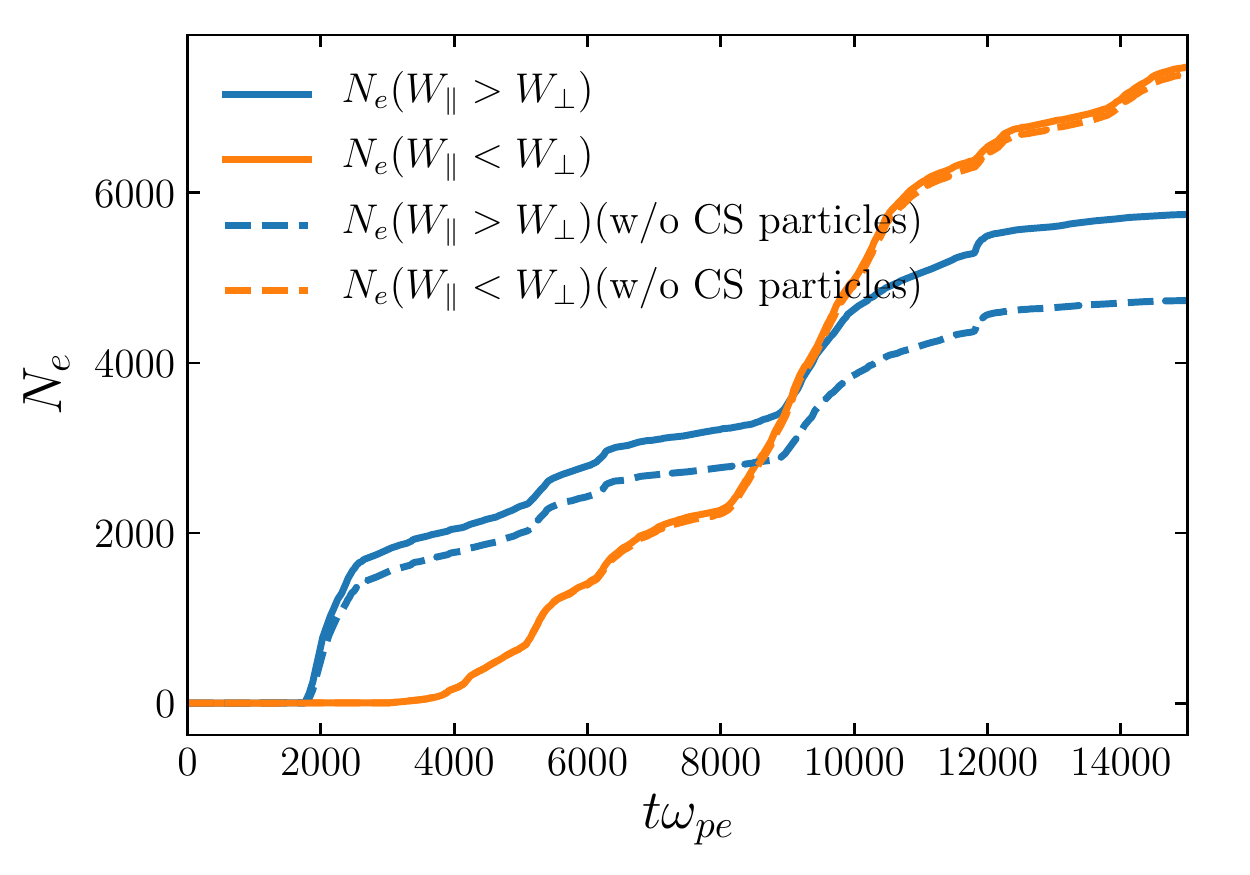}

    \caption{Number of tracer particles that have been injected up to time $t$ and that see a stronger contribution of $W_\parallel$ ($W_\parallel > W_\perp$) or $W_\perp$ ($W_\perp > W_\parallel$) before injection as a function of time $t$.}
    \label{fig:num_wpara_wperp}
\end{figure}

Instead of subjectively classifying trajectories from a limited number of particle trajectories, we resort to statistical quantities computed from all tracer particles. To separate the initial acceleration (``injection'') from later particle acceleration (power-law range) we adopt the injection threshold $\gamma_\mathrm{inj} = \sigma_e / 2$ from \cite{Ball_2019}. For Fig.~\ref{fig:num_wpara_wperp} we check in every time step if a tracer has crossed this threshold for the first time and if so we classify it according to the relative contribution of $W_\parallel$ and $W_\perp$ up to this time. All particles that have exceeded $\gamma_\mathrm{inj}$ at least once by time $t$ are included in the plot of $N_e(t)$. If a particle falls below the threshold it is not removed from the plot. Neither does its classification change if it is re-accelerated and crosses the threshold again. Note that this disadvantages the perpendicular field that tends to act later than the parallel component of the electric field. We have also repeated the analysis by removing the contribution from particles initially in the current sheet and confirmed that they do not modify our conclusion. Plotting the number of particles crossing $\gamma_\mathrm{inj}$ due to more contribution by the parallel electric field ($W_\parallel > W_\perp$) or perpendicular electric field ($W_\perp > W_\parallel$) as a function of time reveals that both $W_\parallel$ and $W_\perp$ contribute to the injection process.
Fig.~\ref{fig:num_wpara_wperp} shows that even in a triggered reconnection setup it takes a while for particles to cross $\gamma_{inj}$ even if particles initially in the current sheet are not excluded.  
The first particles that reached the threshold energy do so due to a dominant contribution from $W_\parallel$.
The time delay and the number of particles that first cross $\gamma_\mathrm{inj}$ likely depend on the details of reconnection onset. Probing this initial phase in a self-consistent way is difficult and requires knowledge about current sheet formation in the specific astrophysical context.
A larger simulation shows a smaller initial jump due to $W_\parallel$ and a smoother increase of both curves over time.
Later in time, particles typically cross the threshold during episodes of plasmoid mergers. This is especially noticeable from time $t \approx 8500\omega_{pe}^{-1}$ when two large plasmoid mergers start as depicted in Fig.~\ref{fig:absJ}. Earlier plasmoid mergers at times $(4500, 5500, 6800)~\omega_{pe}^{-1}$ follow the same trend. Notice that both the number of particle that reach the threshold with $W_\parallel > W_\perp$ and $W_\parallel < W_\perp$ jump at the same time.

\begin{figure}[htb]
 \begin{center}
  \includegraphics[width=\columnwidth]{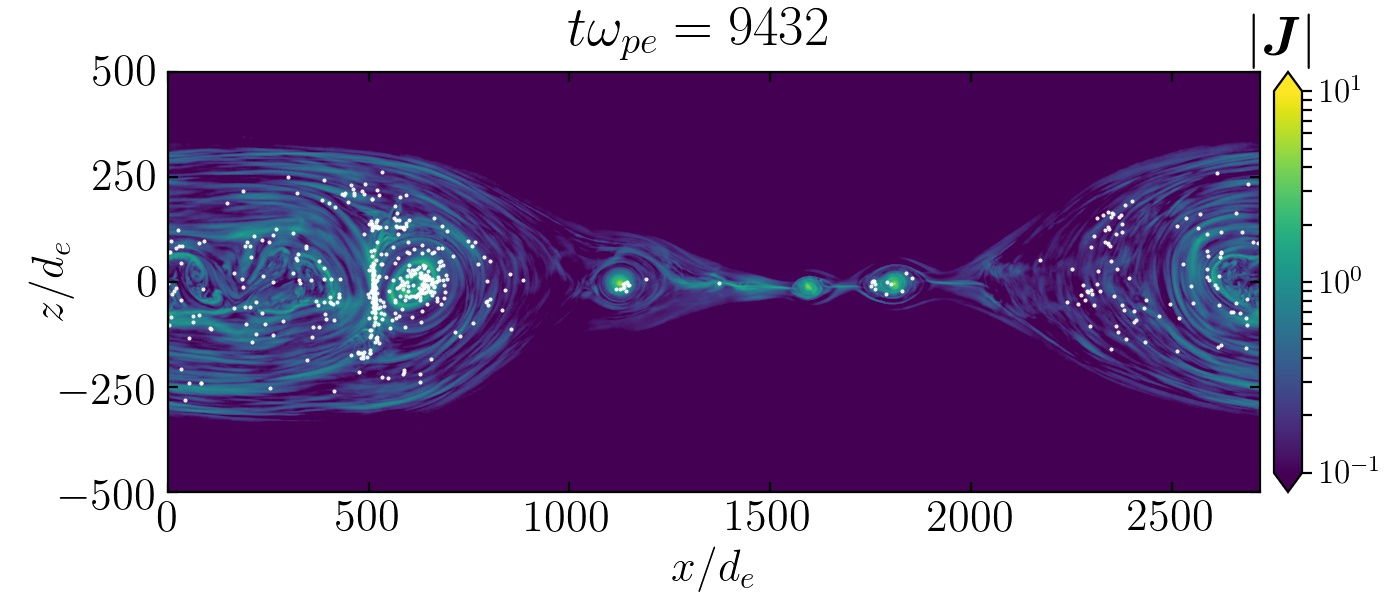}
  \caption{Current density and particle location at $t = 9432\,\omega_{pe}^{-1}$ for particles that cross $\gamma_{inj} = \sigma_e / 2$ around that time.}
  \label{fig:overlay_121}
 \end{center}
\end{figure}

Fig.~\ref{fig:overlay_121} selects all tracer particles that crossed the threshold $\gamma_{inj} = \sigma_e /2$ in the time interval $9393 < t\,\omega_{pe} < 9471$ and shows their location at the middle of the interval on top of the current density at that point in time. Plotting the location of particles as they cross the threshold reveals two preferred locations: The first is secondary current sheets that are formed as islands merge. Strong non-ideal fields are expected as these locations. We still see particles clustering there late in the simulation at $t \sim 10400\,\omega_{pe}^{-1}$ when injection is preferentially done by parallel fields. The other location is the inside of islands as they merge, probably due to the compression of those interacting islands \citep{Li_2018a,Du_2018}.
At late times more injected particles are due to $W_\perp > W_\parallel$. Note that this effect is not visible at early times. The effect of $W_\perp$ is suppressed for another simulation with half the box size and gets more pronounced when doubling the box size. This is examined and discussed in Appendix A. This might indicate that effect of $W_\perp$ is underestimated in simulations with smaller domains.  However even with only two spatially resolved dimensions the simulations are so expensive that further scale up to a box size relevant to astrophysics is not possible.

\begin{figure}[htb]
    \includegraphics[width=\columnwidth]{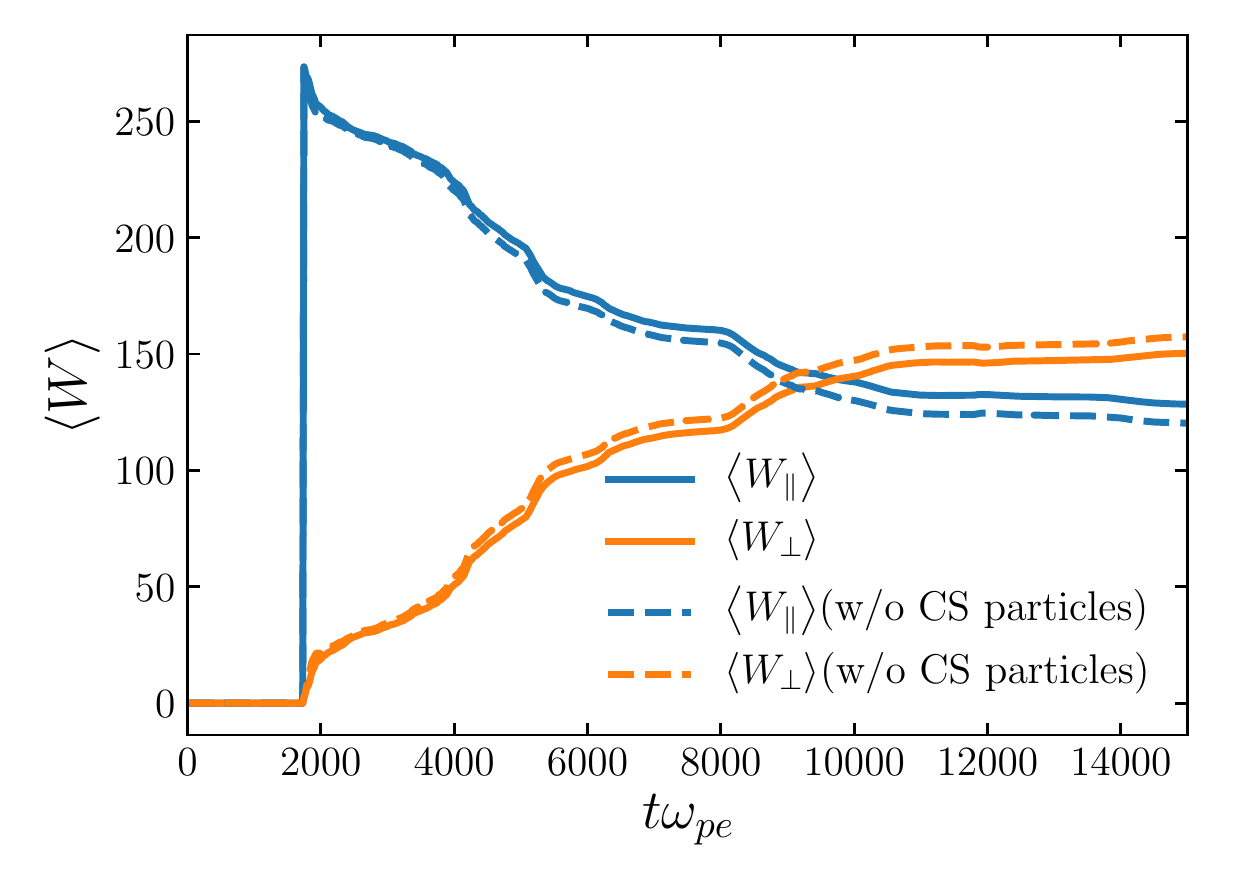}
    
  \includegraphics[width=\columnwidth]{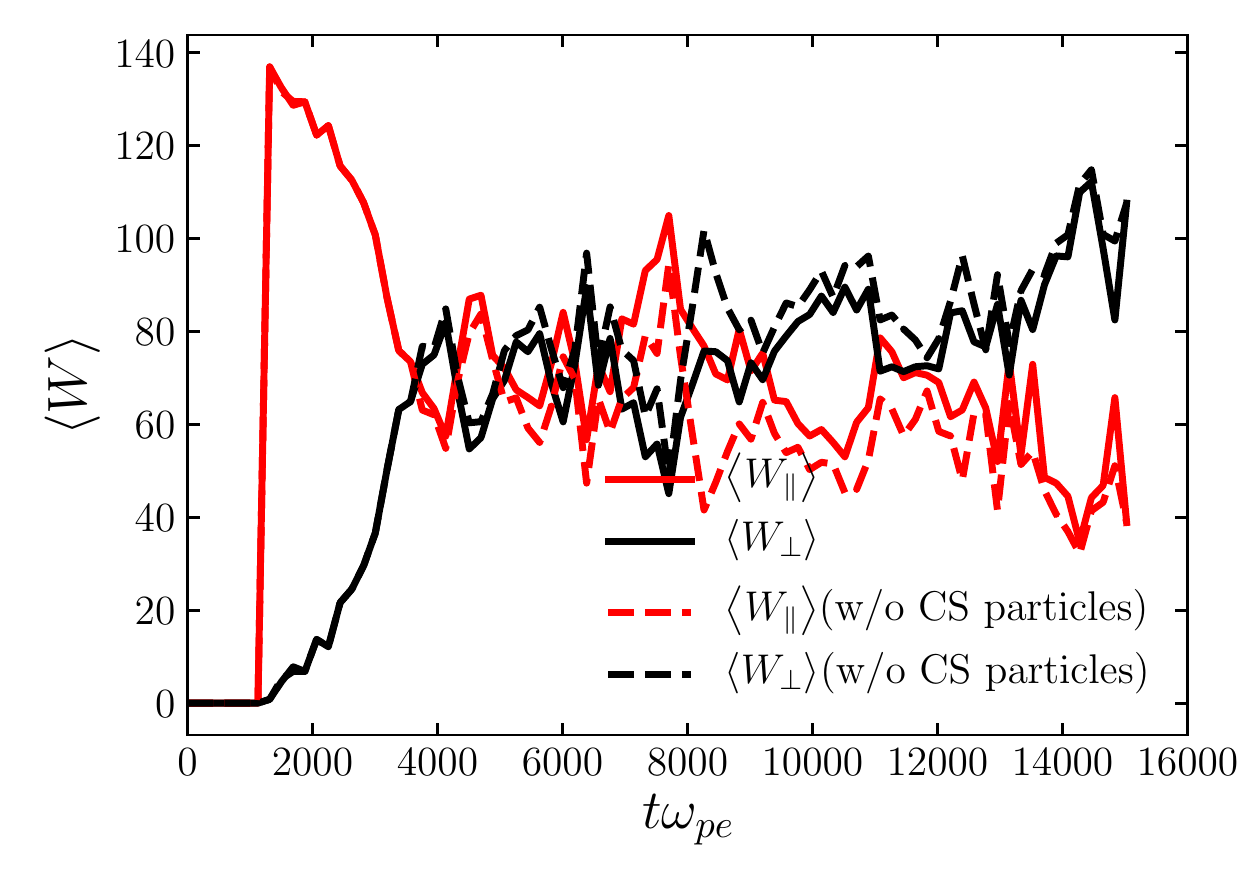}
    \caption{
    Contributions of the work done by the parallel and perpendicular electric fields to particle injection. Top panel: $W_\parallel$ and $W_\perp$ are averaged over all tracer particles that have crossed the injection threshold by time $t$. Bottom panel: $W_\parallel$ and $W_\perp$ averaged over all tracer particles that have crossed $\gamma_{inj} = \sigma_e / 2$ within the last 100 time steps before time $t$.}
    \label{fig:ene_wpara_wperp}
\end{figure}

Fig.~\ref{fig:num_wpara_wperp} counts the number of injected particles due to dominant $W_\perp$ and $W_\parallel$ but does not include the magnitude of  $W_\parallel$ and $W_\perp$. To get a measure of the relative contribution of the two, we average both quantities over all tracer particles at the point in time when the tracer particle crosses $\gamma_\mathrm{inj}$. Note two thing about this plot: the averaging $\left<\dots\right>$ is done over injected particles not over time $t$. Figures plotted with blue/orange colors show quantities calculated for all particles that are injected by time $t$ using quantities at their injection time $t_{inj} \leq t$. This also implies that $W_\parallel + W_\perp = \gamma_\mathrm{inj}$. Fig.~\ref{fig:ene_wpara_wperp}a shows the  resulting averages.
Fig.~\ref{fig:ene_wpara_wperp}b shows how the averages change in time when only considering particles that cross in the last 100 time steps. As in other plots we use red/black for instantaneous quantities. The results are much more noisy due to the limited number of particles available for averaging, but they support the same conclusions.

The initial delay is again visible as well as the fact that the first handful of particles reach $\gamma = \sigma_e /2$ due to $W_\parallel$. Acceleration by the parallel electric field is an effect that occurs in the initial phase of reconnection.
As time progresses the influence of the perpendicular fields become more and more visible as additional particles reach the threshold due to work due by the perpendicular field. Late in the simulation, at two Alfv\'en times ($11300\,\omega_{pe}^{-1}$) the split is about 1.21:1 in favour of $W_\perp$ compared to $W_\parallel$ for particles that started outside the current sheet.
For a smaller simulation of half the size the average contribution of $W_\parallel$ is larger than $W_\perp$. At two Alfv\'en times the split is about 1:4.29 for $W_\perp$ compared to $W_\parallel$ for particles that are initially outside the current sheet.
But for a simulation of twice the size the split improves in favour of $W_\perp$ to about 1.44:1 at two Alfv\'en times for particles that start outside the current sheet (see Appendix). For both large simulations equal contributions are reached at the same time $t \approx 10000 \omega_{pe}^{-1}$.

\begin{figure}[htb]
    \includegraphics[width=\columnwidth]{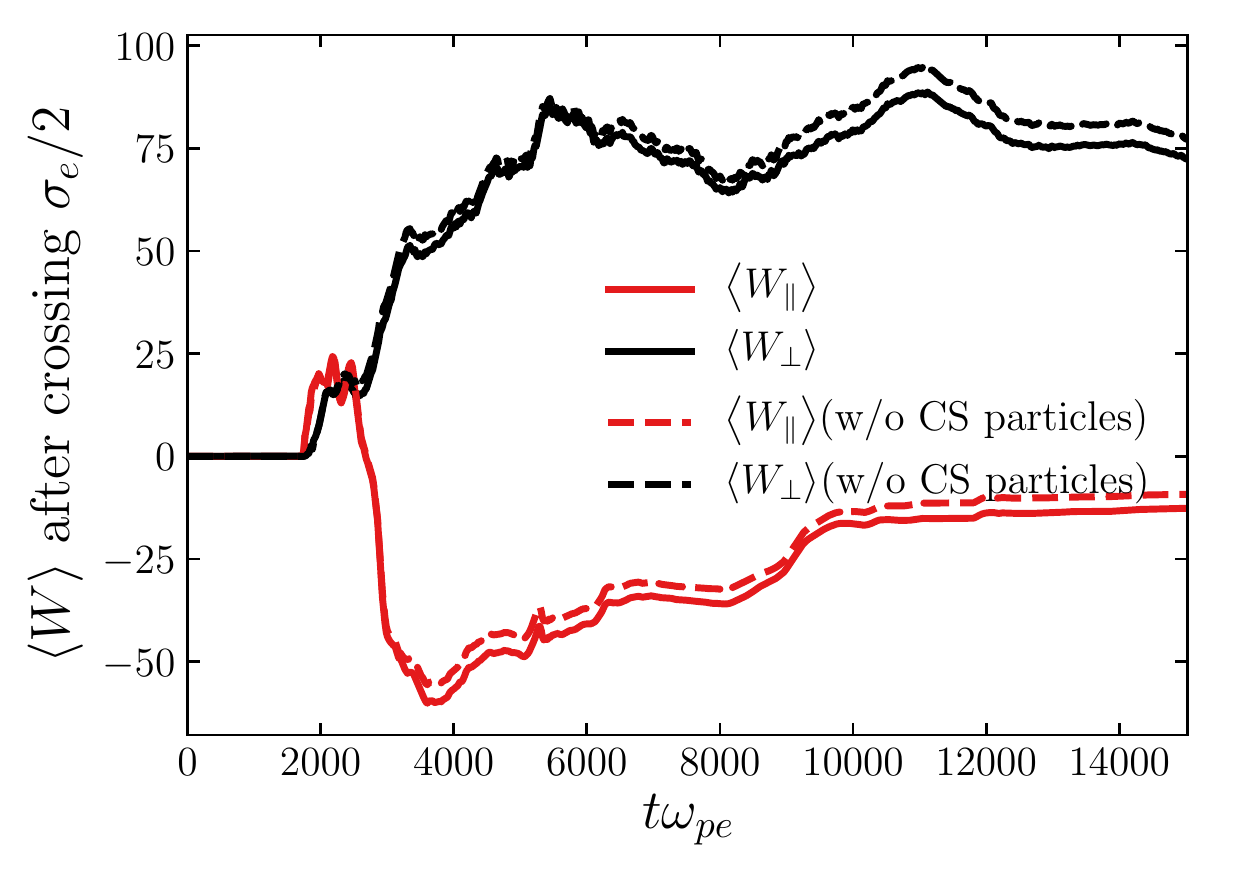}
    \caption{Average energy gain due to parallel and perpendicular electric field of all particles that have already crossed the threshold $\gamma > \sigma /2$ as a function of time. After a very short initial time, energy gain is dominantly due to the perpendicular electric field while the parallel field removes energy from the particles.}
    \label{fig:wpara_wperp_post}
\end{figure}

In addition to the initial energy gain up to $\gamma_{inj}$, we also examine the mechanism of further acceleration that leads to development of the power-law distribution. Fig.~\ref{fig:wpara_wperp_post} shows the averaged energy gain from parallel and perpendicular electric fields of all particles with $\gamma > \gamma_{inj}$ as a function of time. The contribution of the perpendicular electric field dominates over the parallel electric field. In fact, for most of the simulation duration high energy particles lose energy to the parallel electric field and only reach energies significantly above $\gamma_\mathrm{inj}\,m_e\,c^2$ due to $W_\perp$. This fits well with the established picture of secular, late-time energization through a Fermi-type process \citep{Guo_2014,Guo_2019}.

\begin{figure}[htb]
    \includegraphics[width=\columnwidth]{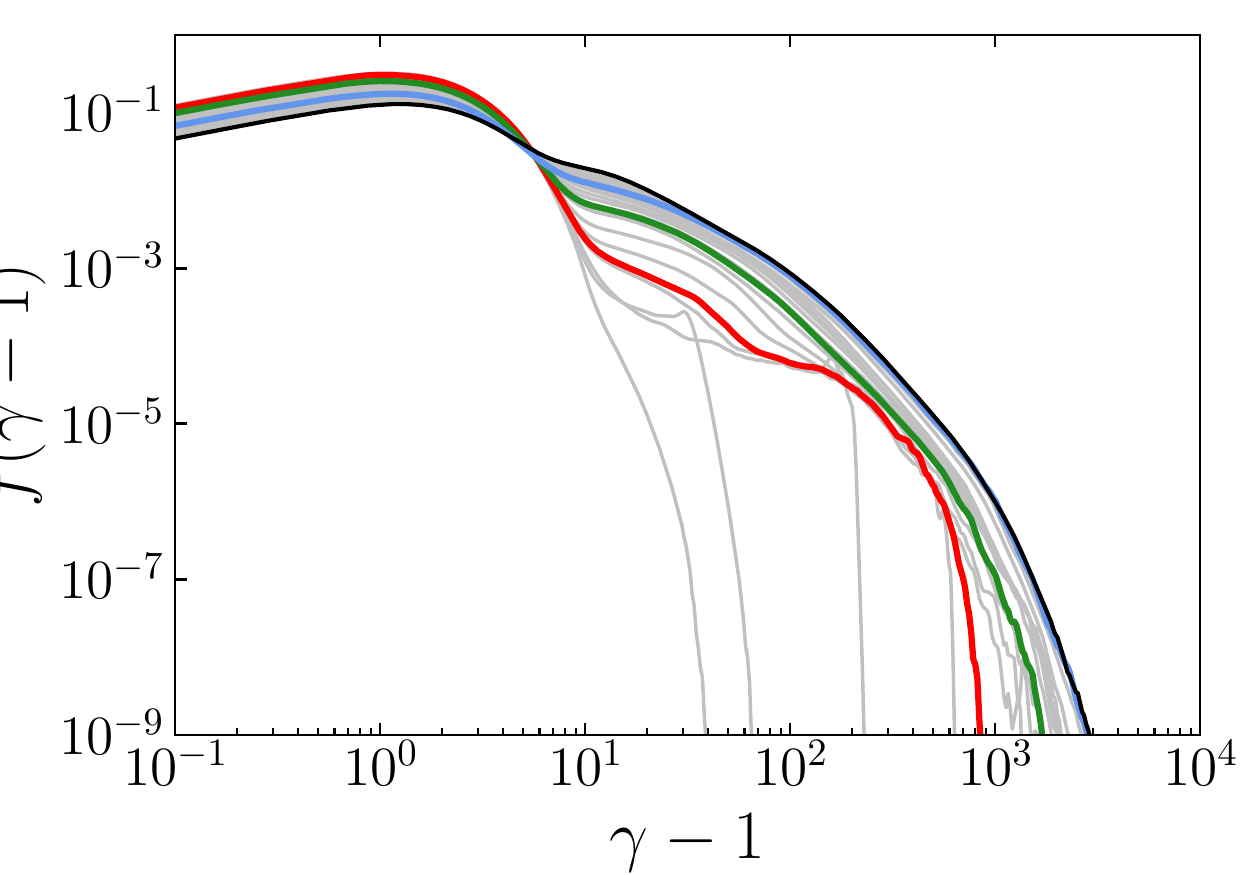}
    \caption{Particle spectra of all tracer electrons at different times. The light gray lines are separated by approximately $780\,\omega_{pe}^{-1}$ and the solid colored lines correspond to the times highlighted in Fig.~\ref{fig:reconnectionrate}. The black line shows the energy spectrum at the end of the simulation.
    \label{fig:spectra}}
\end{figure}

Fig.~\ref{fig:spectra} shows the energy spectra of all tracers at intervals of about $780\,\omega_{pe}^{-1}$. Additionally the spectra at the points in time that are highlighted in Fig.~\ref{fig:reconnectionrate} are plotted. By time $t_1$ when the reconnection rate peaks there are already signs of the heated downstream Maxwellian and a number of non-thermal particles with $\gamma > 100$. A few of them have even reached $\gamma > \sigma_e / 2$ already. At time $t_2$ when the energy conversion rate peaks there are many more particles in the heated downstream. At energies above this heated Maxwellian a powerlaw distribution with $p \approx 2$ has formed. At the late time $t_3$ the downstream Maxwellian and the high-energy tail have grown in particle number. The spectrum has reached its final cutoff at $\gamma \approx 1000$ with a few particles extending up to $\gamma \approx 2000$. In the remaining timestep until the end of the simulation some more particles are processed into heated downstream plasma, but the high-energy tail remains unchanged. This might be an artifact of the 2d simulations performed here that prevents energized particles from accessing energization sites that only appear late in the simulation.

\begin{figure}[htb]
    \includegraphics[width=\columnwidth]{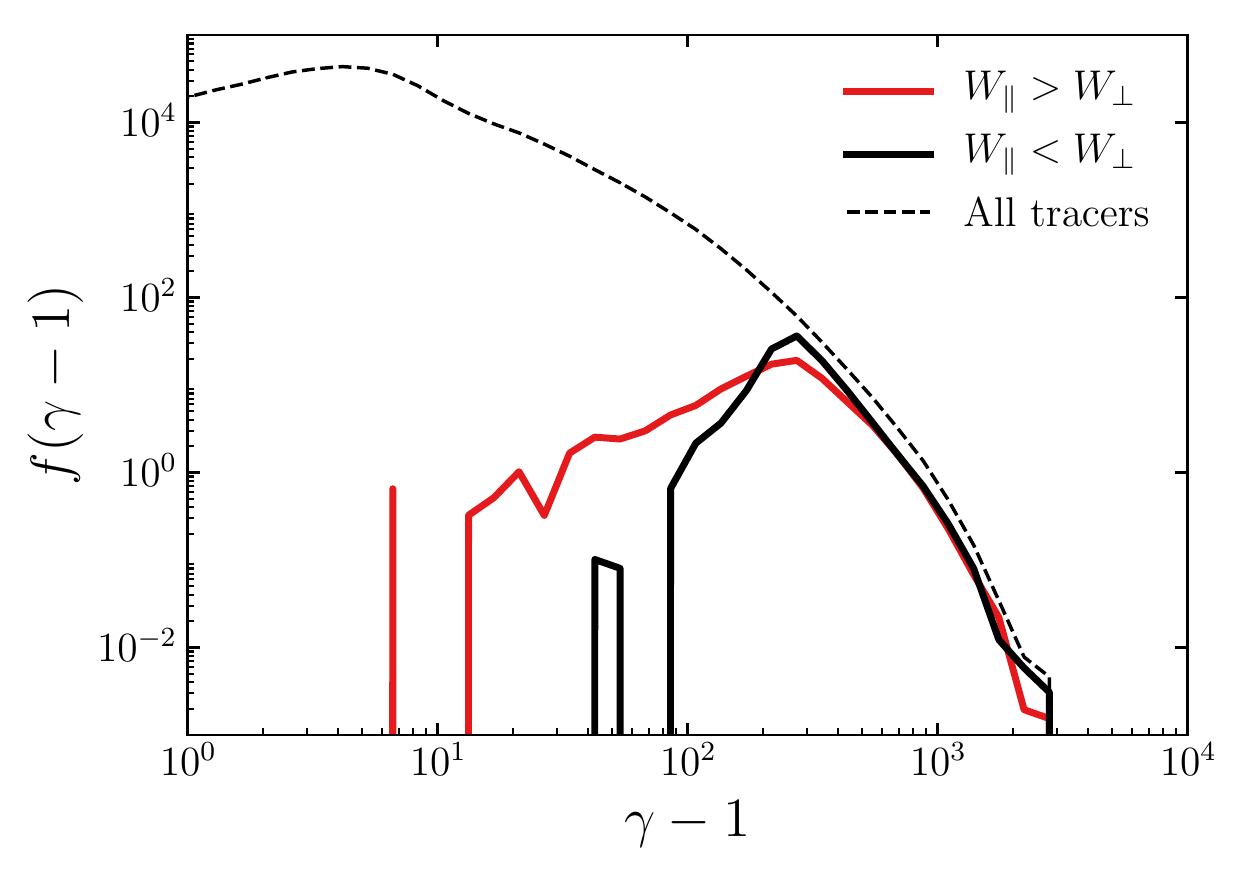}
    \caption{Particle spectrum of electrons that did not start in the current sheet at the end of the simulation, divided according to the work done by the parallel and perpendicular field up to the moment when they cross $\gamma_{inj} = \sigma_e / 2$. Particles that do not appreciably lose energy later in the simulation show very similar spectra. Some particles however do lose their energy at later times, in particular if they originally gain more energy from the parallel field than the perpendicular field.}
    \label{fig:spect_wpara_wperp}
\end{figure}

Fig.~\ref{fig:spect_wpara_wperp} shows the energy spectrum of all tracer particles at the end of the simulation.
Also plotted are the final spectra of the sub-populations that crossed the injection threshold with $W_\parallel > W_\perp$ or $W_\parallel < W_\perp$.  Not all of those particles remain at $\gamma > \sigma_e / 2$ until the end of the simulation, but the ones that do stay in the non-thermal high energy tail exhibit nearly identical spectra, independent of the process that got them across the threshold. This is consistent with all particle acceleration in that energy range coming from $W_\perp$ setting identical spectra, independent of initial source of particle energy below the threshold. Particles that gain energy through $W_\parallel$ initially might be a bit more likely to lose energy again later, but this effect decreases with simulation size and is not significant.

\begin{figure}[htb]
    \includegraphics[width=7cm]{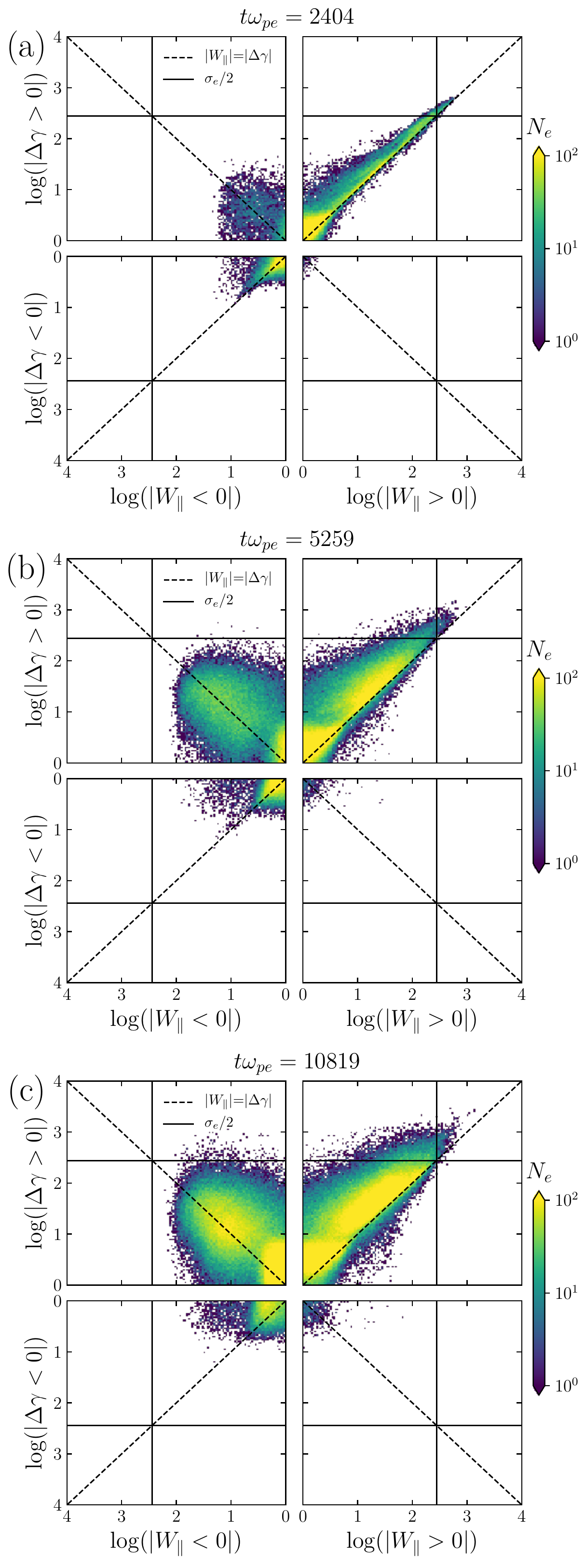}
    \caption{Contribution to energy gain/loss by parallel electric field on all particles that started outside the current sheet.}
    \label{fig:wpara_dgamma_4}
\end{figure}

\begin{figure}[htb]
    \includegraphics[width=7cm]{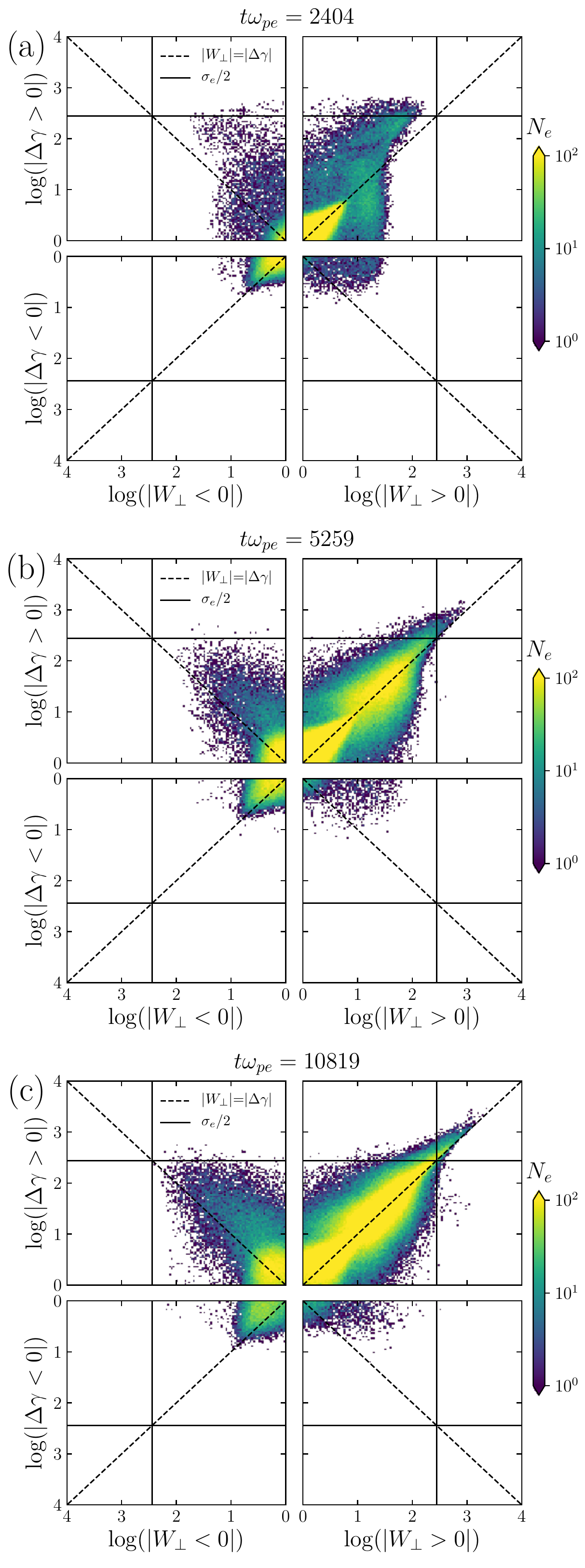}
    \caption{Contribution to energy gain/loss by perpendicular electric field on all particles that started outside the current sheet.}
    \label{fig:wperp_dgamma_4}
\end{figure}

\begin{figure}[htb]
    \includegraphics[width=7cm]{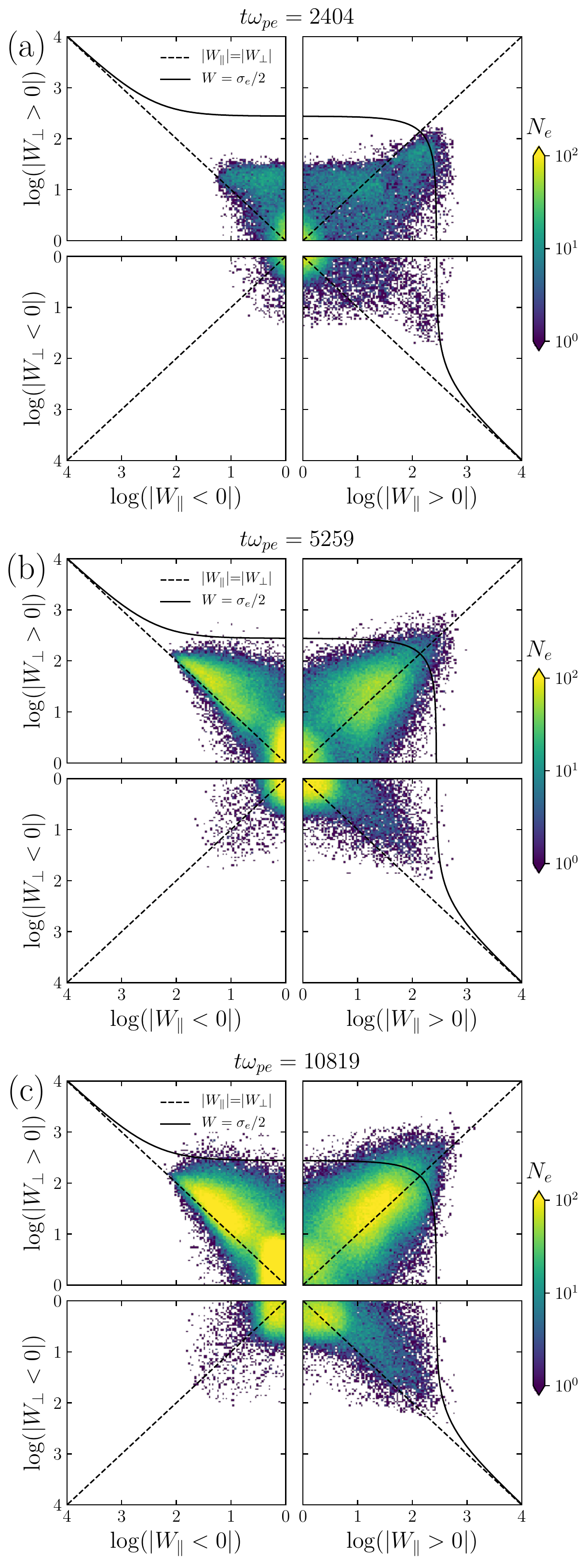}
    \caption{Contribution to energy gain/loss by parallel/perpendicular electric field on all particles that started outside the current sheet.}
    \label{fig:wpara_wperp_4}
\end{figure}

To further investigate the contribution of $W_\parallel$ and $W_\perp$ to $\Delta \gamma$ we looked for a way to visualize the evolution of all three quantities as a function of time for all tracer particles.
Each tracer can be visualized as a point in the three-dimensional $W_\parallel$--$W_\perp$--$\Delta \gamma$ space and moves on the two-dimensional $\Delta \gamma = W_\parallel + W_\perp$ surfaces over time. Fig.~\ref{fig:wpara_dgamma_4} through \ref{fig:wpara_wperp_4} show three projections of this three-dimensional space at times $t_1, t_2$ and $t_3$ as defined in Fig.~\ref{fig:reconnectionrate}. To visualize the density of the point clouds of all 0.6 million tracers we computed histograms that count the number of points in each bin. All three axis of the three-dimensional space can be positive or negative, but span a large range. We therefore decided to plot the negative and positive half of each axis on a separate logarithmic scale. 

Fig.~\ref{fig:wpara_dgamma_4} shows the change in Lorentz factor $\Delta \gamma$ vs the work done by the parallel electric field $W_\parallel$.
The upper right quadrant in each of the three subplots is basically identical to Fig.~10 in \cite{Ball_2019}.
However the other three quadrants also show interesting features.
At early times there is a good correlation between the two quantities as indicated by the bright feature in the histogram along the $\Delta \gamma = W_\parallel$ line. Note however that this feature is mostly located at $\Delta \gamma < \gamma_\mathrm{inj}$.
At later times more particles have reached $\Delta \gamma > \gamma_\mathrm{inj}$, but there is a systematic shift of the peak of the histogram towards $\Delta \gamma > W_\parallel$. This indicated that there is an additional energy gain due to $W_\perp$.
While this shift is visually small (in logarithmic scale), more quantitative analysis in Fig.~\ref{fig:num_wpara_wperp} and Fig.~\ref{fig:overlay_121} have shown the importance of $E_\perp$ during particle injection.

To get a true sense of the the role of the perpendicular electric field it is better to look at Fig.~\ref{fig:wperp_dgamma_4}.
At early times there is a surprising correlation between $\Delta \gamma$ and $W_\perp$ at low energies up to maybe $\gamma \approx 10$. At high energies no such correlation is visible. Many particles have more energy than can be explained by $W_\perp$ alone.
But as time goes on the correlation between $\Delta \gamma$ and $W_\perp$ improves and extends to higher energies. At late times the correlation extend beyond $\sigma_e / 2$ and is better than the correlation with $W_\parallel$. 

Fig.~\ref{fig:wpara_wperp_4} shows the contributions of both parallel and perpendicular electric fields. The resulting change in Lorentz factor is given by their sum.
The threshold $\gamma_\mathrm{inj}$ is indicated in the plot. The curved nature of this line illustrates how deceptive the double logarithmic presentation can be.
At the earliest time $t_1$ particles cross the threshold $\gamma_\mathrm{inj} \approx 275$ indeed mostly due to $W_\parallel$. The majority of high energy particles is however close to the line of equal contributions $W_\parallel \approx W_\perp$.
There is also a large number of particle particles that gained energy dominantly due to $W_\perp$ with energies up to $\gamma \approx 15$.
As time progresses more particles reach high energies, but increasingly close to $W_\parallel \approx W_\perp$. Some particles even cross the threshold with no, or even negative, contribution from $W_\parallel$. On the other hand there are only a few single particles that reach large energies with no (or negative) contributions from $W_\perp$.

\section{Conclusion}
\label{sec:conclusion}

The quest for the origin of power-law energy spectra in magnetic reconnection continues, as more careful analysis reveals more physics insights.
In this paper we performed fully-kinetic simulations of magnetic reconnection
using $\sigma_i = 0.3$ and $\sigma_e = 552.5$. While earlier studies in this trans-relativistic regime ($\sigma_i < 1 < \sigma_e$)  focused on the role of parallel electric field on particle injection (from very low energy to the lower bound $\gamma \sim \sigma_e/2$ of the power-law energy spectra), we study the acceleration by both components of the electric field parallel and perpendicular to the local magnetic field.

We summarize our primary conclusions as below:
\begin{itemize}
    \item The first few particles that reached the injection energy $\gamma \approx \sigma_e /2$ are mostly accelerated by the non-ideal electric field that is parallel to the magnetic field. The acceleration by perpendicular electric field becomes important as the simulation proceeds and eventually outperforms the parallel electric field in terms of particle injection up to the lower energy bound of non-thermal distribution.
    \item The acceleration beyond the low energy ``injection'' to high energy is completely dominated by perpendicular electric field acceleration. The resulting power-law energy spectra, no matter injected primarily by parallel or perpendicular electric field, resemble each other in terms of the spectral index and the high energy break. This provides further support for the Fermi acceleration scenario, as the acceleration to power-law energy is not sensitive to the mechanism and spectral form of the injection processes \citep{Guo_2019}. 
    \item In the trans-relativistic regime, even low-$\beta$ plasmas result in relativistic electron thermal speeds that are sufficient for particles to be picked up by a Fermi-type process\footnote{Sufficiently low plasma $\beta$ will reduce the electron thermal speed to non-relativistic values. This is a regime that has not been considered so far.}. This indicates that the threshold for triggering Fermi acceleration is not a major barrier. 
    In fact, we find that some particles can also be accelerated by a Fermi-like process alone, without a clear preceding acceleration. Of course this process is slower as Fermi acceleration rate scales with the particle energy and is only visible in sufficiently large simulations. While parallel electric field may increase the flux of non-thermal particles by providing a preceding acceleration, perpendicular electric field plays a similar role for particle injection. The energetic particle flux in the simulation will drop significantly without either of the two.
\end{itemize}

We also repeated the analysis shown in this paper for $\gamma_{inj} =\sigma_e / 4$, \textbf{as shown in appendix B,} but this only leads to minor modification of our results. At high energies $W_\perp$ dominates of $W_\parallel$. To get particles up to $\gamma_{inj}$ both $W_\parallel$ and $W_\perp$ can have comparable influence. This is a deviation from the picture that sees $W_\parallel$ acting below $\gamma_{inj}$ and $W_\perp$ above. We conclude that neither of the two should be neglected in the mildly relativistic range where $\gamma$ is between a few and a good fraction of $\sigma_e$. The exact balance of $W_\parallel$ and $W_\perp$ depends on many factors, such as exact choice of $\gamma_{inj}$, system size and guide field strength.
In small systems the role of $W_\parallel$ may be exaggerated. This may be problematic since simulations of astrophysical extend or even the just sufficient size to make robust extrapolations are computationally prohibitively expensive.

Understanding the mechanism of particle injection and further acceleration into a power-law tail allows to infer the particle spectra in realistic astrophysical systems.
Combined with knowledge of the radiation processes it is possible to predict characteristics of the generated radiation such as spectra, polarization and light curves \citep[e.g.,][]{Zhang_2018}.
Comparing prediction with observation results allows to infer the strength, topology and dynamics of the magnetic fields in astrophysical objects that are otherwise difficult to access. 
This is of recent interest in the case of radiatively inefficient accretion disk, such as the disks around Sagittarius A* and M87 that have been observed by the Event Horizon Telescope \citep{Chael_2019}.

\acknowledgments
We gratefully acknowledge discussions with David Ball.
We acknowledge support by the U.S. Department of Energy (DoE) through the Laboratory Directed Research and Development (LDRD) program at Los Alamos National Laboratory (LANL) and DoE/OFES support to LANL, and NASA ATP program through grant NNH17AE68I.
 X.L.'s contribution is in part supported by NASA under grant NNH16AC60I and by NSF/DOE Grant 1902867. F.G.'s  contributions  are in  part  based  upon  work  supported  by  the  U.S.  Department  of  Energy,  Office  of  FusionEnergy Science, under Award Number DE-SC0018240 and DE-SC0020219.
This research was supported by LANL through its Center for Space and Earth Science (CSES). CSES is funded by LANL's LDRD program under project number 20180475DR.
This research used resources provided by the LANL Institutional Computing Program, which is supported by the DoE National Nuclear Security Administration under Contract No. 89233218CNA000001.

%

\bibliography{paper_bib}
\bibliographystyle{aasjournal}

\appendix

\twocolumngrid

\section{Influence of the domain size}

Beside the nominal simulation with a box size of $L_x \times L_z = 2720\,d_e \times 1360\,d_e$ we also performed simulations with other box sizes to examine how domain size influences injection number and averaged energy gain when particle energy crossing $\gamma_{inj} = \sigma_e/2$.

\begin{figure}[htb]
 \begin{center}
  \includegraphics[width=\columnwidth]{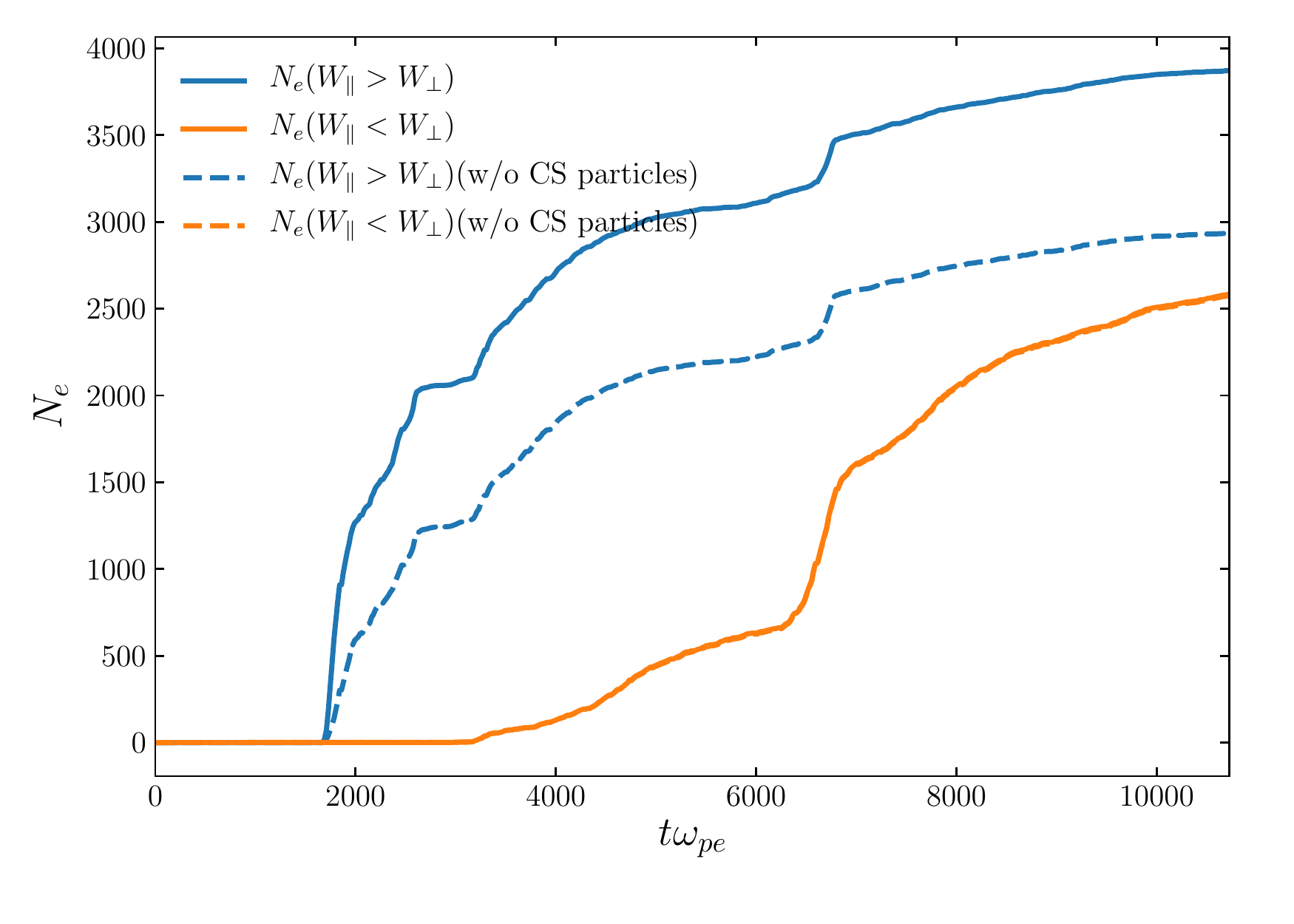}
  \caption{Number of particle dominated by $W_\parallel$ or $W_\perp$ when crossing $\gamma_{inj} = \sigma_e / 2$ in a smaller box of $1360\,d_e \times 680\,d_e$. Analogous to Fig.~\ref{fig:num_wpara_wperp}.}
  \label{fig:num_1360}
 \end{center}
\end{figure}

\begin{figure}[htb]
 \begin{center}
  \includegraphics[width=\columnwidth]{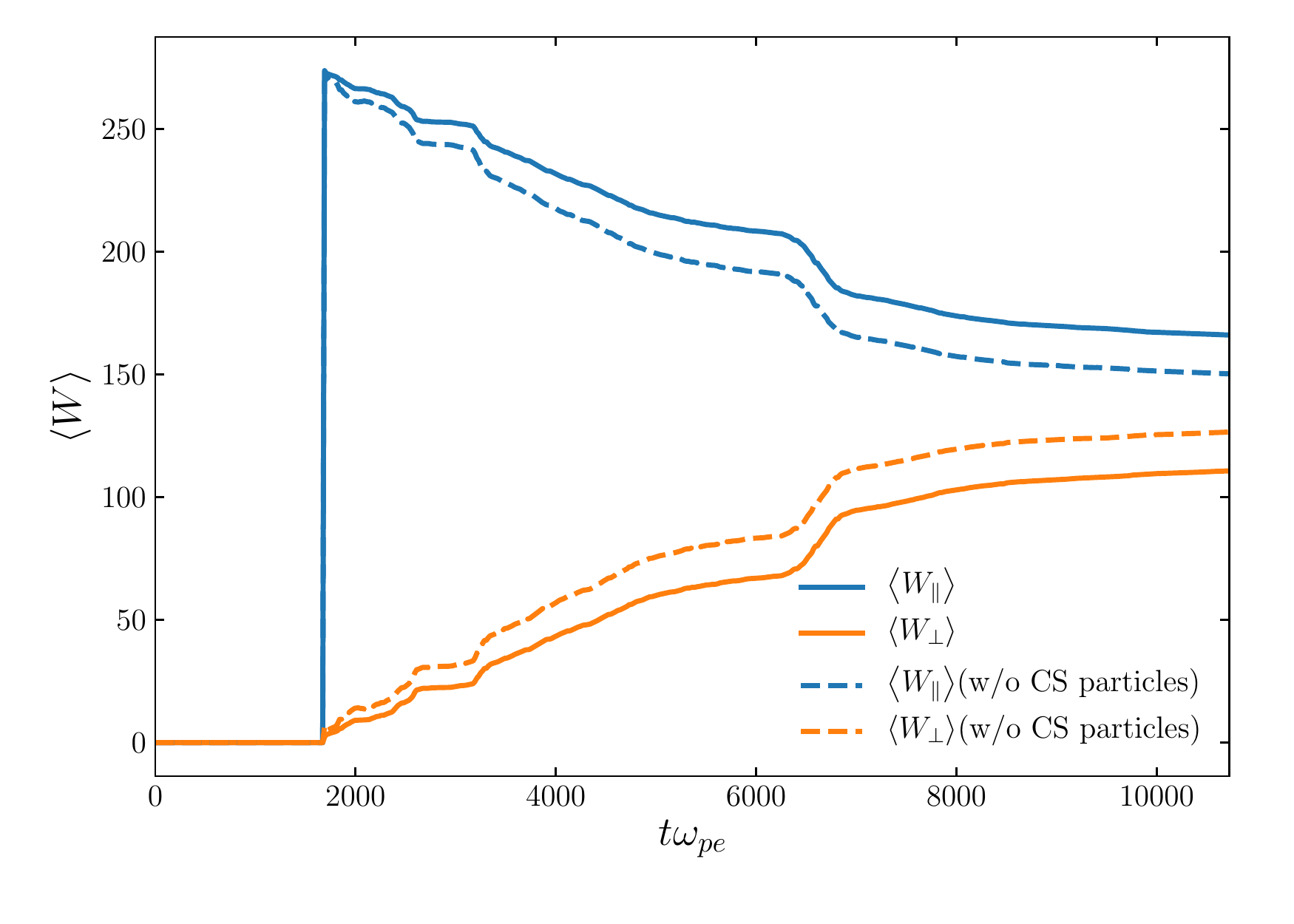}
  \caption{Average energy gain for particles crossing $\gamma_{inj} = \sigma_e / 2$ in a smaller box of $1360\,d_e \times 680\,d_e$, analogous to Fig.~\ref{fig:ene_wpara_wperp}.}
  \label{fig:ene_1360}
 \end{center}
\end{figure}

We performed a \textit{small} simulation with size $L'_x \times L'_z = 1360\,d_e \times 680\,d_e$ using $N'_x \times N'_z = 4096 \times 2048$ grid cells and 167424 tracer particles among the $8.38\cdot10^8$ electrons. Results from this smaller domain size are shown in Fig.~\ref{fig:num_1360} and Fig.~\ref{fig:ene_1360}. Compare to the nominal simulation, a smaller box size leads to an enhancement of $E_\parallel$ acceleration and a reduction of influence of $E_\perp$.

\begin{figure}[ht!]
 \begin{center}
  \includegraphics[width=\columnwidth]{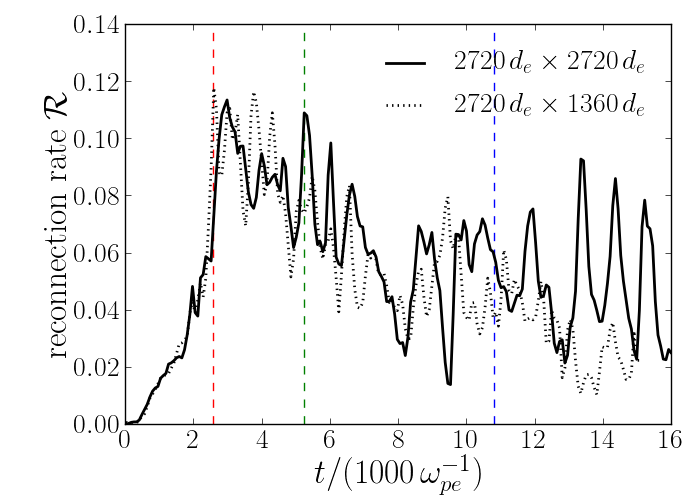}
  \caption{Reconnection rate for $L'_z = 2720\,d_e$. The reconnection rate for the reference simulation that is shown in Fig.~\ref{fig:reconnectionrate} is included as a dotted line.}
  \label{fig:recrate_2720tall}
 \end{center}
\end{figure}

\begin{figure}[ht!]
 \begin{center}
  \includegraphics[width=\columnwidth]{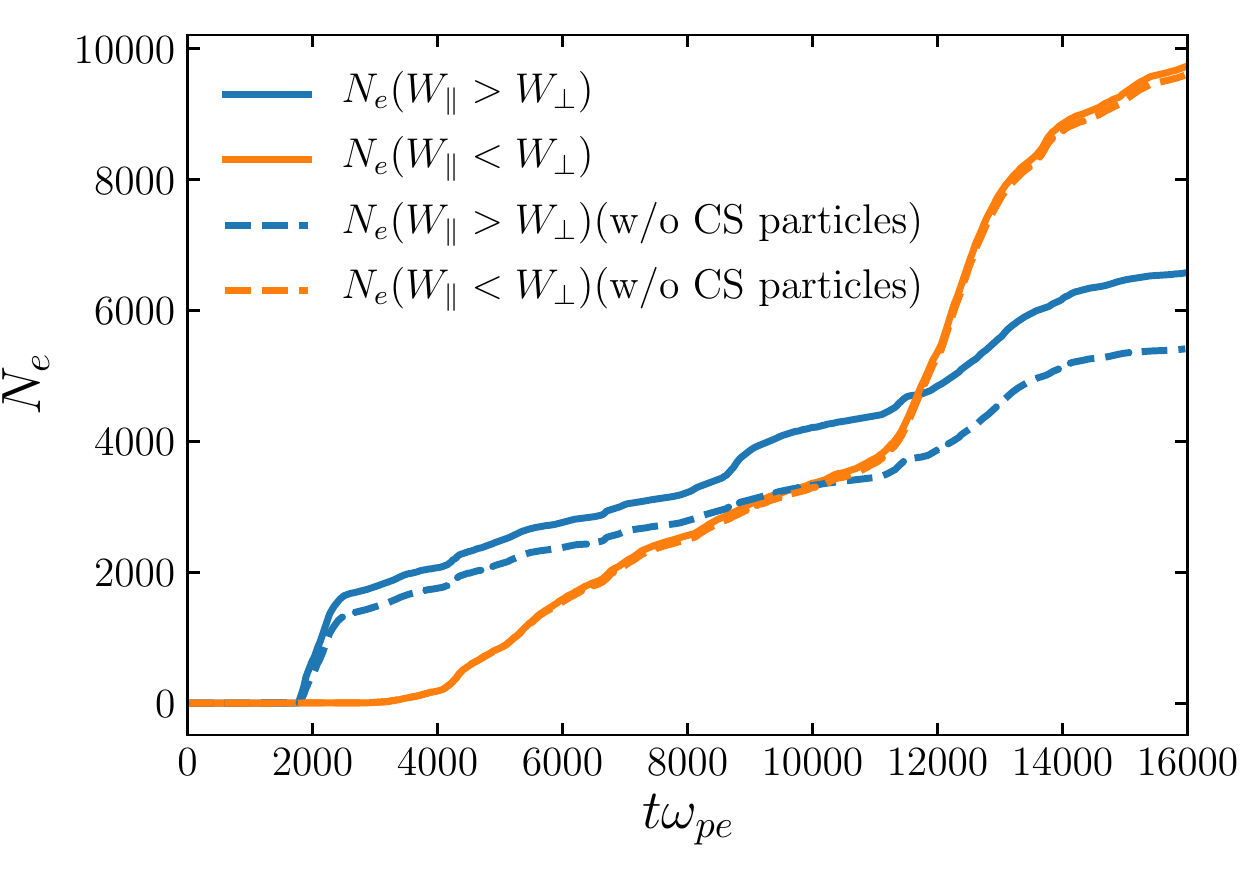}
  \caption{Number of particles crossing $\gamma_{inj} = \sigma_e / 2$ for $L'_z = 2720\,d_e$, analogous to Fig.~\ref{fig:num_wpara_wperp}.}
  \label{fig:num_wpara_wperp_2720tall}
 \end{center}
\end{figure}

\begin{figure}[ht!]
 \begin{center}
  \includegraphics[width=\columnwidth]{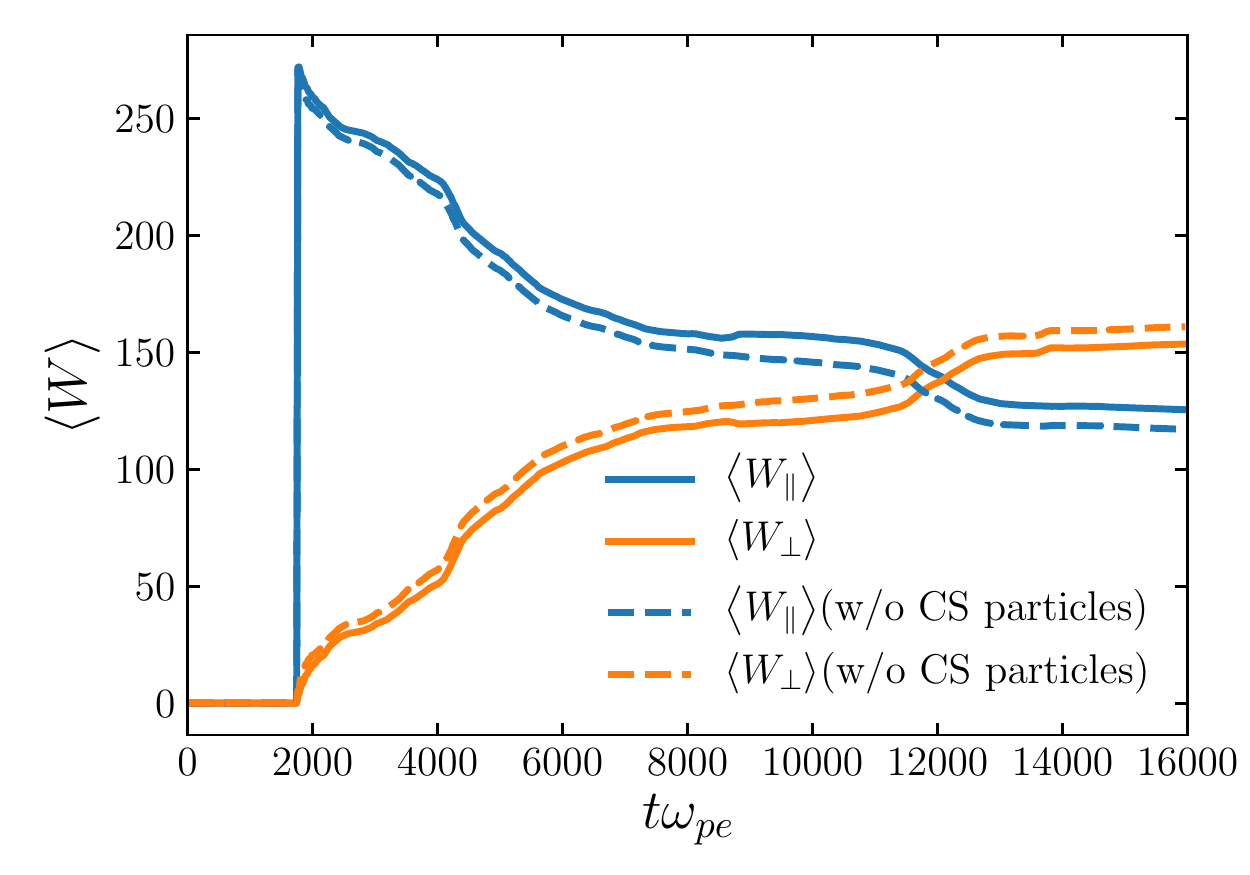}
  \caption{Average energy gain for particles crossing $\gamma_{inj} = \sigma_e / 2$ for $L'_z = 2720\,d_e$, analogous to Figure~\ref{fig:ene_wpara_wperp}a.}
  \label{fig:ene_wpara_wperp_2720tall}
 \end{center}
\end{figure}

We also performed a \textit{square} simulation with size $L'_x \times L'_z = 2720\,d_e \times 2720\,d_e$ using $N'_x \times N'_z = 8192 \times 8192$ grid cells to check for the influence of the upstream boundary conditions. \textbf{There is a small but noticeable difference in the reconnection rate in particular at late times as shown in Fig.~\ref{fig:recrate_2720tall}}. The number of particles that are injected due to $W_\parallel$ and $W_\perp$ also changes somewhat \textbf{as shown in Fig.~\ref{fig:num_wpara_wperp_2720tall}}. \textbf{The average energy gain for particles crossing $\gamma_{inj} = \sigma_e / 2$ is slightly changed (compare Fig.~\ref{fig:ene_wpara_wperp}a with Fig.~\ref{fig:ene_wpara_wperp_2720tall}). Overall}, our main results and conclusions remain unchanged. Because eventually the reconnection outflows still interact with each other via the periodic boundary conditions in the $x$ direction which reduces reconnection and shuts down particle acceleration, we only show our simulation results until about two Alfv\'en crossing time. 

\begin{figure}[htb]
 \begin{center}
  \includegraphics[width=\columnwidth]{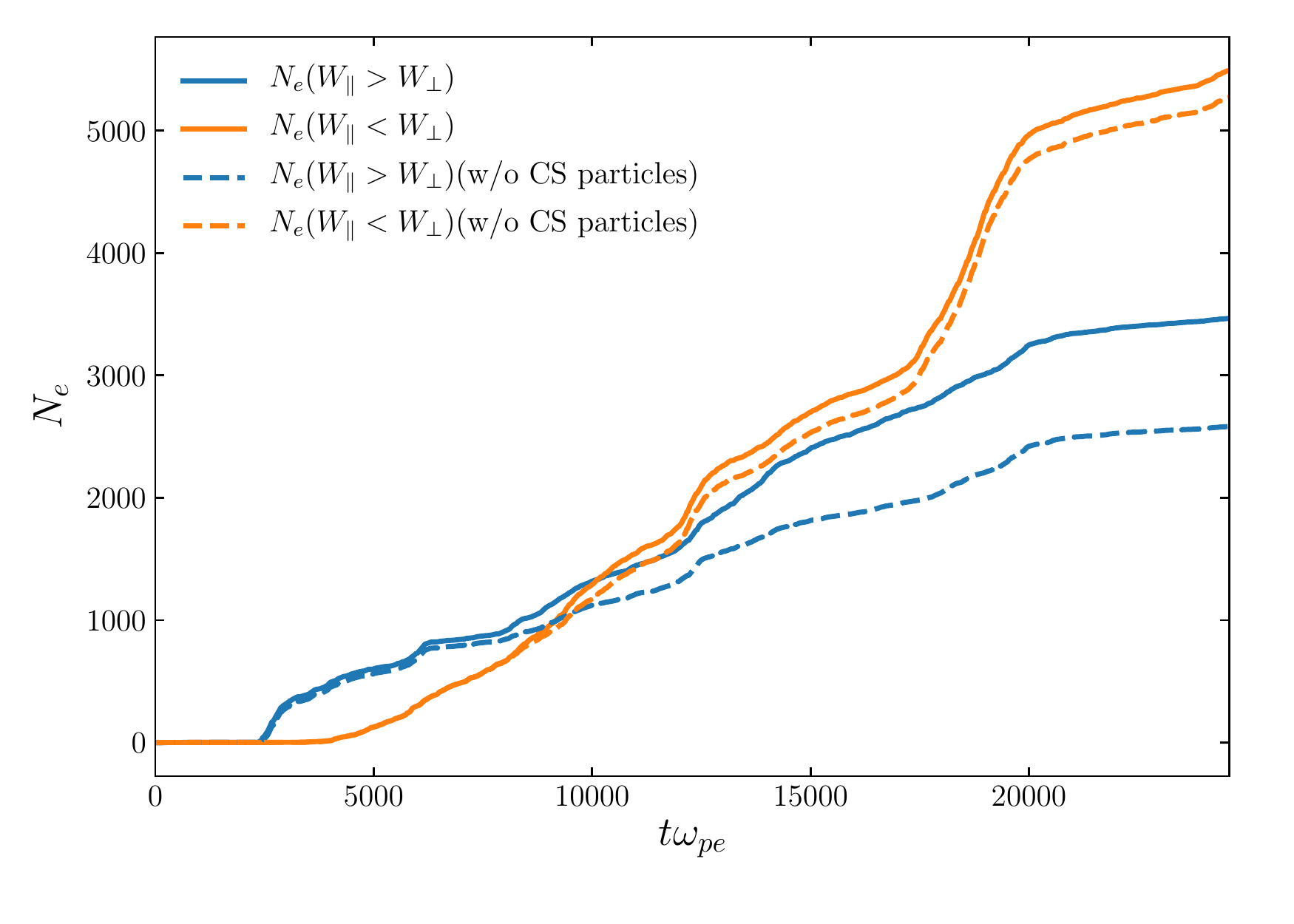}
  \caption{Number of particle dominated by $W_\parallel$ or $W_\perp$ when crossing $\gamma_{inj} = \sigma_e / 2$ in a larger box of $5440\,d_e \times 2720\,d_e$. Compare with Fig.~\ref{fig:num_wpara_wperp} and \ref{fig:num_1360}.}
  \label{fig:num_5440}
 \end{center}
\end{figure}

\begin{figure}[htb]
 \begin{center}
  \includegraphics[width=\columnwidth]{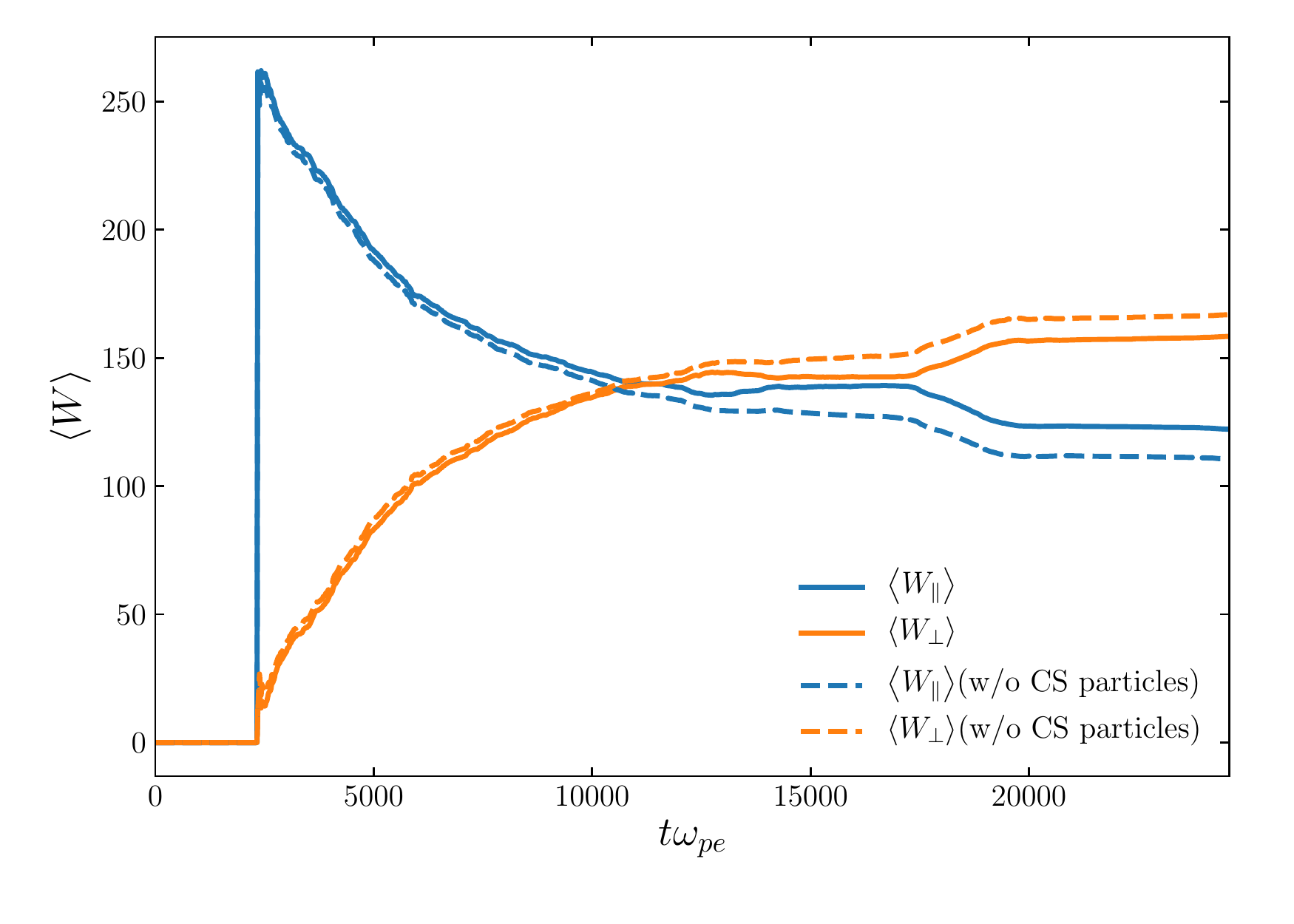}
  \caption{Average energy gain for particles crossing $\gamma_{inj} = \sigma_e / 2$ in a larger box of $5440\,d_e \times 2720\,d_e$. Compare with Fig.~\ref{fig:ene_wpara_wperp} and \ref{fig:ene_1360}.}
  \label{fig:ene_5440}
 \end{center}
\end{figure}

Finally we performed a \textit{large} simulation size $L"_x \times L"_z = 5440\,d_e \times 2720\,d_e$ using $N'_x \times N'_z = 16384 \times 8192$ grid cells and 655360 tracer particles among the $1.34\cdot10^{10}$ electrons.
This larger simulation shows an increased influence of $E_\perp$. From these simulations we conclude that larger simulation domain and longer simulation time lead to more important effects for particle injection through $E_\perp$. 

\section{Influence of the injection threshold}

\begin{figure}[ht!]
 \begin{center}
  \includegraphics[width=\columnwidth]{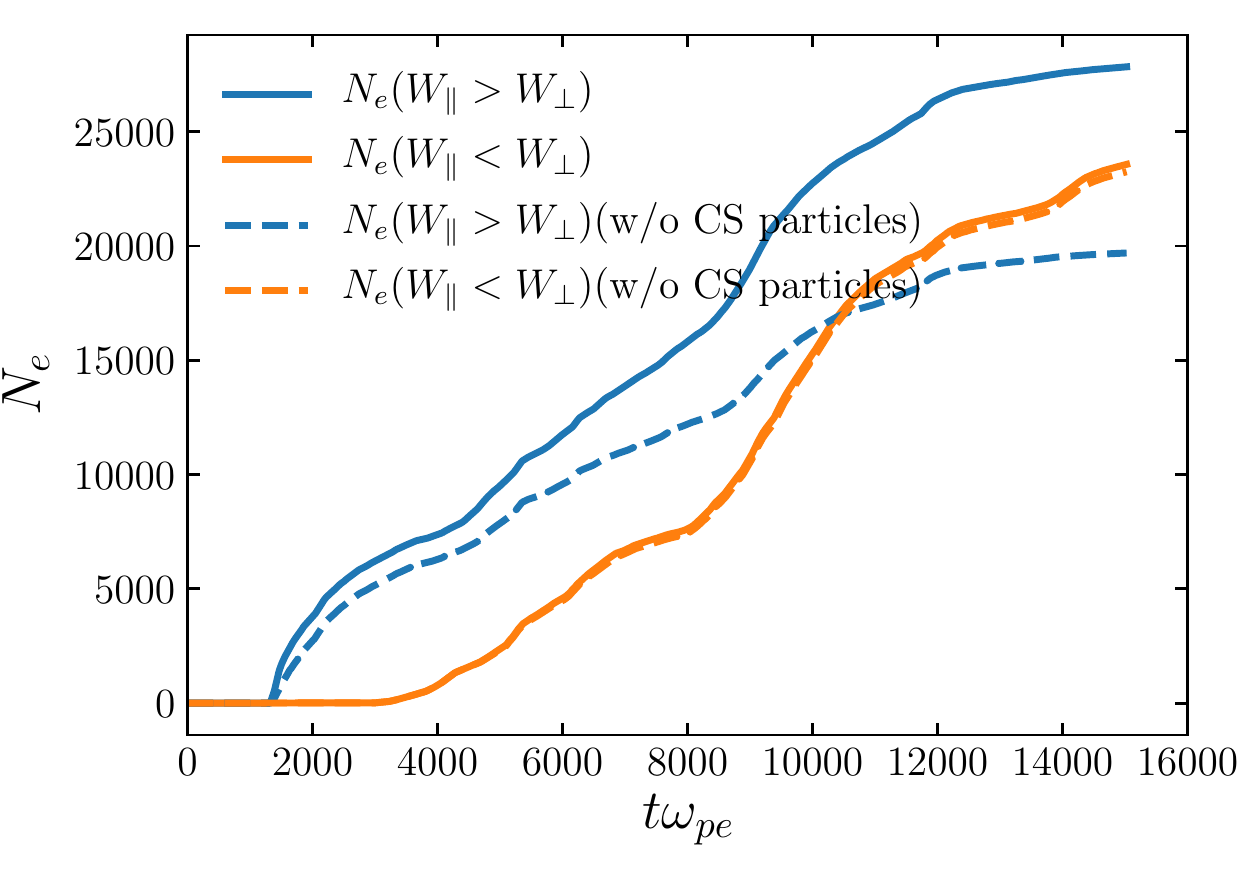}
  \caption{Number of particles crossing the lowered threshold $\gamma_{inj} = \sigma_e / 4$, analogous to Fig.~\ref{fig:num_wpara_wperp}.}
  \label{fig:gamma100_num}
 \end{center}
\end{figure}

\begin{figure}[ht!]
 \begin{center}
  \includegraphics[width=\columnwidth]{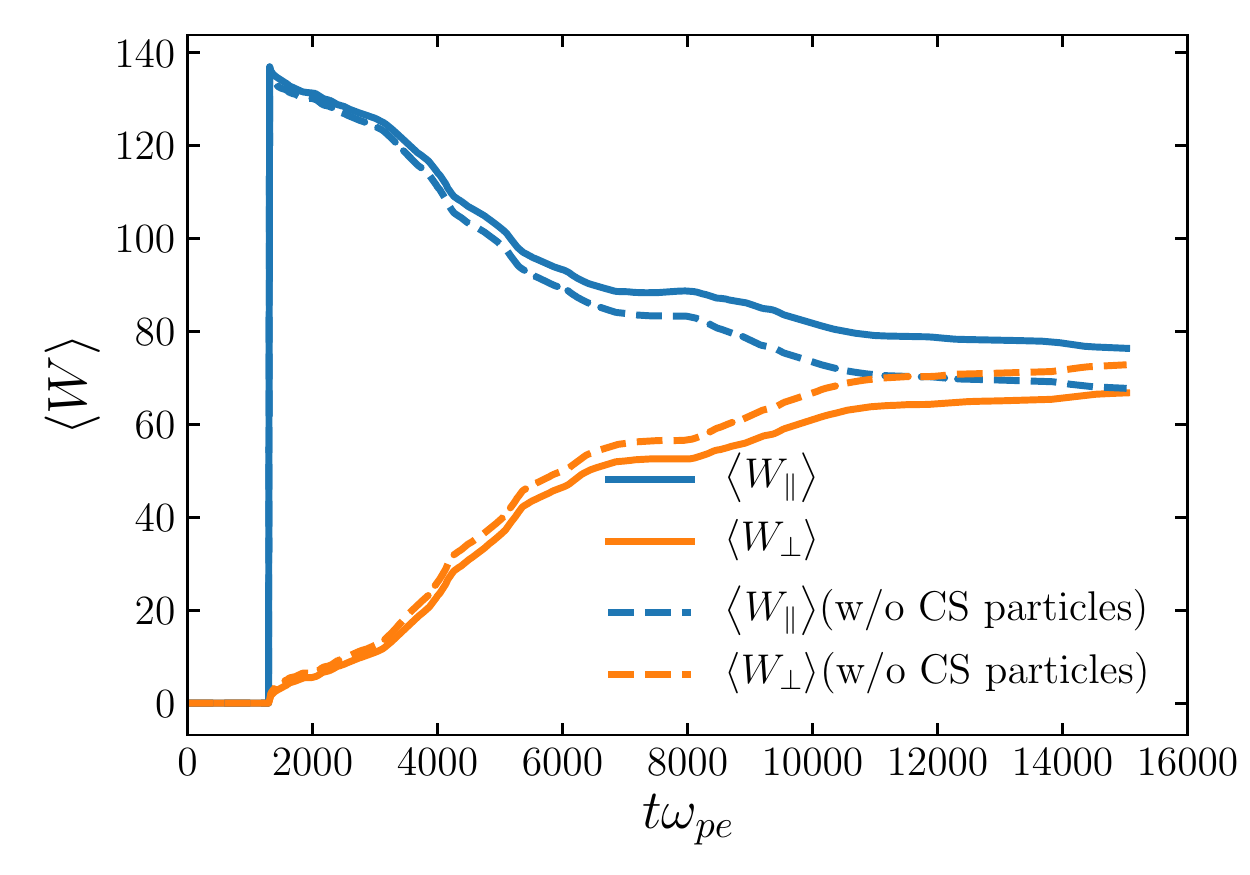}
  \caption{Average energy gain of particles crossing the lowered threshold $\gamma_{inj} = \sigma_e / 4$, analogous to Fig.~\ref{fig:ene_wpara_wperp}a.}
  \label{fig:gamma100_ene}
 \end{center}
\end{figure}

\textbf{We also repeated the analysis shown in the main text for $\gamma_{inj} =\sigma_e / 4$. In this appendix we show the two most important plots for the simulation with a box size of $L_x \times L_z = 2720\,d_e \times 1360\,d_e$ from that analysis. Comparing Figure~\ref{fig:gamma100_num} with Fig.~\ref{fig:num_wpara_wperp}, we see that still injection by $W_\parallel > W_\perp$ is the first process that quickly picks up particles. The time until the first particles cross the lowered $\gamma_{inj}$ is reduced a little bit. The total number of particles that go over the threshold until the end of the simulation goes up by about a factor of 4. The fact that $W_\parallel > W_\perp$ is important for particles that start inside the current sheet whereas $W_\perp > W_{\parallel}$ nearly exclusively acts on particles that start outside the current sheet is more pronounced. This leads to the conclusion that for particles outside the current sheet $W_\perp$ is still more important than $W_\parallel$. For particles that start inside the current sheet this is no longer true. Note however that the particles that start inside the current sheet are heavily affected by the choice of the initial equilibrium (Harris current sheet vs force-free current sheet) and by the fact that we start with a current sheet that is thin enough to be unstable instead of waiting for the current sheet to thin down dynamically. We therefore are reluctant to make strong claims about those particle. This is consistent with the analysis in \cite{Ball_2018}, where the authors removed those particles from their analysis.}

\textbf{Compared to Fig.~\ref{fig:ene_wpara_wperp}a in the main text, the relative contribution of $W_\perp$ is slightly reduced compared to $W_\parallel$ when the injection threshold is lowered to $\gamma_{inj} = \sigma_{e}/4$ (shown in Figure~\ref{fig:gamma100_ene}). For particles that start outside the current sheet it is still (barely) dominant. For earlier times or particles that start inside the current sheet it is slightly smaller. However in all cases except very early times it is of comparable magnitude and can not simply be neglected.}

\end{document}